\documentclass[useAMS,usenatbib,referee]{biom}

\usepackage{algorithm}
\usepackage{algorithmicx}
\usepackage{algpseudocode}
\usepackage{amsmath,epsfig,amssymb,bm,amsfonts}
\usepackage{xcolor,graphicx}
\usepackage{hyperref,subfigure, multirow, setspace,  booktabs}

\graphicspath{{figures/}}
\newcommand{\nvs}{\vspace{-0.26in}}

\newtheorem{Theorem}{Theorem}
\newtheorem{Pro}{Proposition}

\def\bse{\begin{eqnarray*}}
\def\ese{\end{eqnarray*}}
\def\be{\begin{eqnarray}}
\def\ee{\end{eqnarray}}
\def\bsq{\begin{equation*}}
\def\esq{\end{equation*}}
\def\bq{\begin{equation}}
\def\eq{\end{equation}}

\def\wh{\widehat}

\def\rank{\mbox{rank}}

\def\bb{{\boldsymbol\beta}}

\def\bg{\boldsymbol\gamma}

\def\btheta{\boldsymbol\theta}

\def\bLambda{\boldsymbol \Lambda}

\def\A{{\bf A}}
\def\U{{\bf U}}
\def\V{{\bf V}}

\def\B{{\bf B}}
\def\E{{\bf E}}

\def\D{{\bf D}}
\def\V{{\bf V}}

\def\h{{\bf h}}
\def\b{{\bf b}}
\def\I{{\bf I}}
\def\M{\mbox{ $\mathcal{M}$}}
\def\M{{\bf M}}

\def\F{{\bf F}}
\def\H{{\bf H}}

\def\bSigma{{\bf \Sigma}}

\def\U{{\bf U}}
\def\S{{\bf S}}

\def\W{{\bf W}}

\def\X{{\bf X}}
\def\S{{\bf S}}
\def\x{{\bf x}}
\def\I{{\bf I}}

\def\Y{{\bf Y}}

\def\0{{\bf 0}}
\def\Z{{\bf Z}}
\def\z{{\bf z}}

\def\bq{\begin{equation}}
\def\eq{\end{equation}}

\def\wh{\widehat}

\def\trans{^{\rm T}}

\def\squarebox#1{\hbox to #1{\hfill\vbox to #1{\vfill}}}
\def\btheta{{\boldsymbol \theta}}
\def\balpha{{\boldsymbol \alpha}}

\def\bse{\begin{eqnarray*}}
\def\ese{\end{eqnarray*}}
\def\be{\begin{eqnarray}}
\def\ee{\end{eqnarray}}
\def\bsq{\begin{equation*}}
\def\esq{\end{equation*}}
\def\bq{\begin{equation}}
\def\eq{\end{equation}}
\def\wh{\widehat}

\def\trans{{\mathrm{T}}}
\def\boxit#1{\vbox{\hrule\hbox{\vrule\kern6pt\vbox{\kern6pt#1\kern6pt}\kern6pt\vrule}\hrule}}

\def\red{\color{black}}

\title{High-Dimensional Covariate-Augmented Overdispersed  Poisson  Factor Model}
\author{Wei Liu$^{1}$, Qingzhi Zhong$^2\ast$ \\
$^1$Cardiovascular and Metabolic Disorders Program,\\ Duke-NUS Medical School, Singapore \\
$^2$School of Economics, Jinan University, Guangzhou, China\\
*Corresponding author.   Email: \emph{zhongqz19@icloud.com}
}

\begin{document}

\label{firstpage}

\begin{abstract}
The current Poisson factor models often assume that the factors are unknown, which overlooks the explanatory potential of certain observable covariates. This study focuses on high dimensional settings, where the number of the count response variables and/or covariates can diverge as the sample size increases. A covariate-augmented overdispersed Poisson factor model is proposed to jointly perform a high-dimensional Poisson factor analysis and estimate a large coefficient matrix for overdispersed count data. A group of identifiability conditions are provided to theoretically guarantee computational identifiability. We incorporate the interdependence of both response variables and covariates by imposing a low-rank constraint on the large coefficient matrix. To address the computation challenges posed by nonlinearity, two high-dimensional latent matrices, and the low-rank constraint, we propose a novel variational estimation scheme that combines Laplace and Taylor approximations.  We also develop a criterion based on a singular value ratio to determine the number of factors and the rank of the coefficient matrix. Comprehensive simulation studies demonstrate that the proposed method outperforms the state-of-the-art methods in estimation accuracy and computational efficiency. The practical merit of our method is demonstrated by an application to the CITE-seq dataset. A flexible implementation of our proposed method is available in the R package \emph{COAP}.
\end{abstract}

\maketitle
\section{Introduction}

High-dimensional large-scale count data is becoming increasingly prevalent in the era of big data, spanning from microbial data~\citep{xu2021zero} to single-cell DNA/RNA/protein sequencing data~\citep{mimitou2021scalable} collected by various next-generation sequencing platforms. Although this type of data offers new avenues for research and applications, it also gives rise to  substantial challenges due to its unique characteristics, such as being count type, overdispersion, high dimensionality of variables, and large-scale sample size.  In particular, describing the dependence among variables and extracting meaningful information from this type of data  are formidable tasks. Furthermore, there may be additional high-dimensional covariates that contribute to the variability of the original high-dimensional count variables. For example, our motivating peripheral blood
mononuclear cells dataset, collected by the CITE-seq sequencing platform~\citep{stoeckius2017simultaneous}, comprises a cell-by-gene count matrix with thousands of cells and tens of thousands of genes, along with additional hundreds of protein markers, which can account for gene expression levels.

Factor models as a powerful framework for analyzing high-dimensional large-scale data  has gained significant attentions in recent years~\citep{ bai2013principal,liu2023generalized}, and has found extensive applications in different fields~\citep{chen2020structured, bai2013principal, liu2022joint,liu2023probabilistic}.
Most existing methods study high-dimensional linear factor models (LFMs) by assuming that the observed variables are continuous and have a linear relationship with the latent factors \citep{fan2017sufficient,li2018embracing, chen2021quantile}.
However, LFMs have limitations in analyzing count-valued data as they cannot account for the count nature of the data or utilize the explanatory potential of additional covariates. Poisson factor models, as proposed by \cite{btt091} and \cite{xu2021zero}, offer a viable alternative by modeling the dependence of count data on latent factors via a log-linear model. \cite{liu2023generalized} has also introduced a flexible generalized factor model that can accommodate mixed-type variables, including count variables as a special case. However, none of the above methods can effectively incorporate additional covariates. To address this limitation, {\red \cite{chiquet2018variational} and \cite{hui2017variational} proposed  probabilistic Poisson PCA model  and generalized linear latent variable model (GLLVM), respectively.  However, both  probabilistic Poisson PCA and GLLVM are unsuitable for high-dimensional covariates because they lack structural restrictions on the large coefficient matrix, and the implementation algorithm suffers the intensive computation.}

To overcome the  limitations of existing methods, we propose the COAP model, a \underline{c}ovariate-\underline{a}ugmented \underline{o}verdispersed \underline{P}oisson factor model, that simultaneously accounts for the high-dimensional large-scale count data with high overdispersion and additional high-dimensional covariates. Our model is highly versatile, as it encompasses two commonly used models for count data analysis: Poisson factor models~\citep{btt091} and multi-response Poisson regression models~\citep{yee2003reduced}.  Specifically, we assume that the high-dimensional count variables, also known as response variables in the regression framework, are  modeled  by the high-dimensional covariates and unobservable factors through a log-linear model. Different from \cite{chiquet2018variational}, we introduce an additional error term in the log-linear model to  account for the possible overdispersion not explained by the latent factors. Furthermore, we assume the large regression coefficient matrix to be low-rank, characterizing the interdependency of both high-dimensional count variables and the covariates, and allowing for a valid estimator of the high-dimensional coefficient matrix.  Finally, we develop a computationally efficient algorithm to implement our model.

Our contribution is as follows. First, in contrast with existing models, {\red we introduce a novel model for handling count data.} Second, we theoretically investigate the computational identifiability of the proposed model to uniquely determine the parameters in models, which is crucial for interpretability. Third,  we propose a variational EM (VEM) algorithm to simultaneously estimate the variational parameters, the large coefficient matrix and loading matrix. To overcome the computational challenges induced by two high-dimensional latent matrices and the nonconjugate model, we develop a novel combination of Laplace and Taylor approximation.  This produces iterative explicit solutions for the complicated variational parameters. Moreover, we apply matrix analysis theory to derive two methods with iterative explicit solutions for the optimization problem of the large coefficient matrix with a rank constraint. The proposed VEM algorithm is computationally efficient, with linear complexity in both sample size and count variable dimension. 
We also theoretically prove the convergence of the VEM algorithm. Fourth, we develop a criterion based on a singular value ratio to determine the number of factors and the rank of coefficient matrix under the COAP framework. {\red Finally, numerical experiments show that COAP outperforms existing methods in estimation accuracy and computational efficiency.} 

The rest of the paper is organized as follows.  We introduce the model setup of COAP and study its identifiability conditions in Section
\ref{sec:model}. In Section
\ref{sec:est}, we present the estimation method, the variational EM algorithm of COAP, and the tuning parameter selection based on a singular value ratio criterion.  
The performance of COAP is assessed through simulation studies and real data analysis in Section
\ref{sec:simu} and \ref{sec:real}, respectively.  A brief
discussion about further research in this direction is given in
Section \ref{sec:dis}.  Technical proofs are relegated to {\red the
Supporting Information.}
\nvs
\section{Method} \label{sec:model}
\subsection{Model setup}

To explicate the COAP model setup, we use the motivating  CITE-seq dataset as an example. The dataset consists of a count matrix $\X=(\X_1,\cdots,\X_p) \in \mathbb{R}^{n\times p}$, where $\X_j=(x_{1j},\cdots,x_{nj})^{\trans}$, $n$ is the number of cells and $p$ is the number of genes. Each entry $x_{ij}$ corresponds to the expression level of gene $j$ in cell $i$. In addition to the count matrix, protein marker expressions, denoted by $\z_i \in \mathbb{R}^d$, are available as covariates to account for the gene expression levels. Notably, our model allows $p$ and $d$ to diverge as $n$ increases. Furthermore, $p$ can be much greater than $n$ while $d$ is less than $n$. To account for the count nature, high overdispersion, and high-dimensional covariates, we assume that the gene expression $x_{ij}$ depends on both the covariates $\z_i$ and a group of latent factors $\h_i\in \mathbb{R}^q$, with a conditional distribution that follows an overdispersed Poisson distribution:
\vspace{-0.2in}
\begin{eqnarray}
  &&x_{ij}|\widetilde   y_{ij} \sim  Pois(a_i \widetilde   y_{ij}), \label{eq:xij}\\
  && y_{ij} = \ln (\widetilde   y_{ij}) = \z_i^\trans\bb_j + \h_i^\trans\b_j + \varepsilon_{ij}, \label{eq:yij}
\end{eqnarray}
where $a_i$ is the normalization factor for the $i$th cell, the unobservable Poisson rate $\widetilde   y_{ij}$ and its logarithm $y_{ij}$ represent the underlying expression and log-expression level of gene $j$ in cell $i$, respectively. These Poisson rates depend on the covariates $\z_i$ and the latent factors $\h_i$ via a log-linear model. The latent factor $\h_i$ accounts for both unknown or unavailable covariates from $\z_i$,  and correlation among genes.
The $d$-vector  $\bb_j$ includes an intercept representing the mean log-expression of the gene $j$, and $d-1$ regression coefficients for covariates $\z_i$. Let $\bb = (\bb_1, \cdots, \bb_p)^{\trans}$ and $\mathrm{rank}(\bb)=r \ll \min\{p, d\}$.
$\b_j$ is the loading vector in the factor model, and $\varepsilon_{ij}$ is an error term such that $\varepsilon_{ij} \stackrel{i.i.d.}\sim N(0,\varsigma_j)$. The key differences {\red among COAP and the  probabilistic Poisson PCA and GLVVM models} are the low-rank structure of $\bb$ and the extra $\varepsilon_{ij}$ in COAP. These enable COAP to handle high-dimensional count responses and covariates, and explain possible overdispersion not accounted for by the factors. 
Additionally, in the context of CITE-seq data and microbiome data, the scalar $a_i$ is typically set to represent the sequencing depth or to one, depending on whether the analysis focuses on absolute or relative levels, as discussed in previous studies~\citep{sun2020statistical, kenney2021poisson}.

\nvs
\subsection{Model identifiability conditions}
Computational identifiability  is usually studied in factor model to make the estimate of factor and loading matrix unique~\citep{bai2013principal}.
Models \eqref{eq:xij} and \eqref{eq:yij} are unidentifiable due to two facts: (a) $\h_i^\trans \b_j= \h_{i,new}^\trans\b_{j,new}$ holds by setting {$\h_{i,new}=\M^{\trans}\h_i$ and $\b_{j,new}=\M^{-1}\b_j$ for any invertible matrix $\M$};
and (b) $\h_i$ can be projected on $\z_i$ if they are linearly dependent, i.e., $\h_i={\balpha}\z_i  + \h_{i,new}$ with $\balpha\in \mathbb{R}^{q\times d}$, then $\z_i^\trans\bb_j + \h_i^\trans\b_j=\z_i^\trans \bb_{j,new} + \h_{i,new}^\trans\b_j$ with $\bb_{j,new}=\bb_j+\balpha^\trans\b_j$. Fact (a) is intrinsic in factor models~\citep{bai2013principal} while fact (b) is unique in our considered model, introduced  by the augmented covariates.

Let $\I_q$ be a $q$-by-$q$ identity matrix. Denote $\H=(\h_1,\cdots,\h_n)^\trans,\B=(\b_1,\cdots,\b_p)^\trans,\Z=(\z_1,\cdots,\z_n)^\trans$ and $\boldsymbol\varsigma=(\varsigma_1, \cdots, \varsigma_p)^{\trans}$. To guarantee the identifiability of the proposed model, we propose a group of computational identifiability conditions to uniquely determine the  parameters in models \eqref{eq:xij} and \eqref{eq:yij}. We impose three conditions:
\vspace{-0.2in}
\begin{description}
  \item[(A1)] $\frac{1}{n}\H^\trans\H = \I_q$; 
  \item[(A2)] $\B^\trans\B$ is diagonal with decreasing diagonal elements and the first nonzero element in each column of $\B$ is positive;
  \item[(A3)] The column space of $\H$ is linearly independent of that of $\Z$ and $\rank(\Z)=d$.
\end{description}
\vspace{-0.2in}
Conditions (A1) and (A2)  are commonly used to resolve the unidentifiability resulting from fact (a) in factor models \citep{bai2013principal,liu2023generalized}. Condition (A3) states that the extracted factor $\H$ accounts for the variability in the count data that cannot be explained by the covariates $\Z$, and is necessary to address the unidentifiability of fact (b). Condition (A3) is also crucial in ensuring that our models fully and parsimoniously capture the data structure. 

We state the identifability in the following theorem, its proof is deferred to
Supporting Information.
\nvs
\begin{Theorem}
 Under Conditions (A1)--(A3), models \eqref{eq:xij} and \eqref{eq:yij} are computationally identifiable.
\end{Theorem}

\nvs

\section{Estimation}\label{sec:est}
Denote $\Y$ an $n\times p$ matrix with $(i,j)$-th entry $y_{ij}$. {\red If $(\Y,\H)$ is observable, all parameters can be estimated by maximizing the conditional log-likelihood function of $(\X, \Y)$ given $(\Z,\H)$:
\nvs
\begin{eqnarray} \label{eq:logFullP}
  \ln P(\X, \Y|\Z,\H) &=& \ln P(\X|\Y) + \ln P(\Y|\H,\Z) \nonumber\\
    &=& \sum_{i,j} \left[x_{ij}y_{ij} - a_i\exp(y_{ij}) - \frac{1}{2}\{(y_{ij}-\z_i^\trans \bb_j - \b_j^\trans\h_i)^2/\varsigma_{j} + \ln \varsigma_{j} \} \right],
\end{eqnarray}
after dropping a constant.} However, $\Y$ and $\H$ are unobservable and high-dimensional, thus, it is very difficult to maximize \eqref{eq:logFullP}. {\red Similar to \cite{liu2023generalized} and \cite{wang2022maximum}, we treat $\H$ as a parameter and  treat $\Y$ as latent variables}, and invoke the EM algorithm \citep{dempster1977maximum}. The E-step of EM algorithm requires to compute the posterior distribution of latent variables, {\red $P(\Y|\X,\Z,\H)$}. Since \eqref{eq:logFullP} involves high-dimensional random matrices $(\Y,\X,\Z)$ and the exponential term $\exp(y_{ij})$, it is untractable to compute the posterior distribution based on Bayes' formula. We propose a variational EM algorithm to transfer the computation of {\red $P(\Y|\X,\Z,\H)$} to the estimation of variational parameters. To achieve this, we employ a mean field variational family~\citep{blei2017variational} {\red $q(\Y)$} to approximate $P(\Y|\X,\Z,\H)$, and seek the optimal approximation  that minimizes the KL divergence of {\red $q(\Y)$} and $P(\Y|\X,\Z,\H)$, where
\begin{eqnarray}\label{q_approx}
\red q(\Y)= \Pi_{i=1}^n\Pi_{j=1}^p N(y_{ij}; \mu_{ij}, \sigma^2_{ij}),
\end{eqnarray}
and $N(y; \mu, \sigma^2)$ is the normal density function of $y$ with mean $\mu$ and variance $\sigma^2$. Thus, the E-step and M-step can be formulated to maximize the evidence lower bound (ELBO) with respect to (w.r.t.) $(\btheta,\bg)$: {\red
\begin{eqnarray}
  ELBO(\btheta,\bg) &=& \mathrm{E}_q  \ln P(\X, \Y|\Z, \H) - \mathrm{E}_q \ln q(\Y) \nonumber\\
  &=& \sum_i \sum_j \bigg\{x_{ij}\mu_{ij} -  a_i\exp(\mu_{ij}+\sigma_{ij}^2/2) - \frac{1}{2}\{(\mu_{ij}-\z_i^\trans \bb_j -\b_j^\trans\h_i)^2/\varsigma_{j} + \varsigma_{j}^{-1}\sigma_{ij}^2  \nonumber\\
  && + \ln \varsigma_{j} \} \bigg\}  + \frac{1}{2}\sum_i\bigg\{\sum_j \ln (\sigma_{ij}^2)\bigg\} + c , \label{eq:elbo}
\end{eqnarray}
where $\btheta=(\bb,\B, \H, \boldsymbol\varsigma)$ is the model parameter, and ${\bg} = (\mu_{ij},\sigma^2_{ij}, 1\leq i\leq n,1\leq j\leq p)$ is the variational parameter,} $\mathrm{E}_q$ represents taking an expectation w.r.t.  $\Y$ based on the density function $q(\Y)$, and $c$ is a general constant that is independent of $(\btheta,\bg)$.
\nvs
\subsection{Variational EM algorithm}
\subsubsection{Variational E-step}
First, we consider to update the variational parameters of $\Y$.  Because $q(\Y)$ is not a conjugate distribution to $P(\X|\Y)$, a direct maximization of \eqref{eq:elbo} w.r.t.  $(\mu_{ij}, \sigma^2_{ij})$ is very difficult. We turn to use a Laplace approximation~\citep{wang2013variational} for $q(\Y)$. Specifically, by the fact that $P(\Y|\X,\Z) \propto P(\X|\Y) P(\Y|\H,\Z)$, we construct a Gaussian proxy for the posterior by adopting the Taylor
approximation around the maximum posterior point of $P(\X|\Y) P(\Y|\H,\Z)$.
A direct calculation yields $\ln \left\{P(\X|\Y) P(\Y|\H,\Z)\right\}= \sum_i \sum_j [x_{ij}y_{ij} - a_i\exp(y_{ij}) - \frac{1}{2}\{(y_{ij}-\z_i^\trans \bb_j - \b_j^\trans\h_i)^2/\varsigma_{j}\}] + c$. 
Denote {\red $f_{ij}(y)=x_{ij}y- a_i\exp(y) - \frac{1}{2}\{(y-\z_i^\trans \bb_j-\b_j^\trans\h_i)^2/\varsigma_{j}\}$},  the posterior mean and variance of $y_{ij}$ can be estimated by
\begin{equation}\label{mu_sigma}
\widehat \mu_{ij} = \arg\max_{y} f_{ij}(y), \widehat \sigma^2_{ij} = -f''_{ij}(\widehat\mu_{ij})^{-1}, i=1,\cdots,n,j=1,\cdots,p,
\end{equation}
where {\red $f''_{ij}(y)={d^2 f_{ij}(y)}/{dy^2}=-a_i\exp(y) - \varsigma_{j}^{-1}$.}

Equation \eqref{mu_sigma} requires numerical optimization to find the maximum for all $\{f_{ij}(y), i=1,\cdots,n, j=1,\cdots, p\}$, which can be computationally expensive for large $n$ and $p$. To improve the computational efficiency, we enhance the Laplace approximation by a novel combination with Taylor's expansion. Let $g_i(y) = - a_i \exp(y)$ and $g_i'(y)={dg_i(y)}/{dy}$. Applying the Taylor's expansion to $g_i(y)$ for $y$ in a small neighborhood of $y_0$, we have
\vspace{-0.2in}
\begin{equation}\label{expan_giy}
g_i(y) \approx g_i(y_0) + g_i'(y_0) (y-y_0) + \frac{1}{2} g_i''(y_0) (y-y_0)^2,
\end{equation}
where $y_0$ is the previous iterative value of $\mu_{ij}$.
Substituting \eqref{expan_giy} into $f_{ij}(y)$, differentiating $f_{ij}(y)$ w.r.t.
$y$ and setting the derivatives to zero lead to:
\begin{equation}\label{eq:updateY}
 \widehat\mu_{ij}=\frac{x_{ij} - a_i \exp(y_0)(1-y_0)+\varsigma_{j}^{-1}( \z_i^\trans \bb_j + \b_j^\trans\h_i)}{\varsigma_{j}^{-1} +a_i \exp(y_0)},  \,\  \widehat\sigma_{ij}^2=\frac{1}{a_i\exp(\widehat\mu_{ij}) + \varsigma_{j}^{-1}}.
\end{equation}
These explicit solutions significantly improve the computational speed.

\subsubsection{Variational M-step}

{\red Let $\bSigma_{\varepsilon\varepsilon}(\boldsymbol\varsigma)=\mathrm{diag}(\varsigma_1, \cdots, \varsigma_p)$. Differentiating $ELBO(\btheta,\bg)$ w.r.t. $\b_j$  and $\h_i$, and set the derivatives to zero, we have
\nvs
\vspace{0.1in}
\begin{eqnarray}
   \b_j &=& (\H^{\trans}\H)^{-1} \sum_i \h_i(\mu_{ij} - \z_i^\trans \bb_j), \label{eq:updata_bj} \\
   \h_i &=& \{\B^{\trans}\bSigma_{\varepsilon\varepsilon}^{-1}(\boldsymbol\varsigma)\B\}^{-1} \sum_j \b_j(\mu_{ij} - \z_i^\trans \bb_j)/\varsigma_{j}. \label{eq:updata_hi}
\end{eqnarray}
}
\vspace{-0.2in}
Next, we update $\bb$ and $\boldsymbol\varsigma$. Taking into account the low-rank structure of $\bb$,
we estimate  $(\bb, \boldsymbol\varsigma)$ by maximizing {\red
\vspace{-0.2in}
\begin{eqnarray}
   && \sum_i \sum_j \left[- \frac{1}{2}\{(\mu_{ij}-\z_i^\trans \bb_j -\b_j^\trans\h_i)^2/\varsigma_{j} + \varsigma_{j}^{-1}\sigma_{ij}^2 + \ln \varsigma_{j} \} \right], \nonumber \\
   & & \mbox{subject to rank}(\bb)=r.   \label{eq:sumBeta}
\end{eqnarray}
Let $\widetilde\Y$ be the $n$-by-$p$ matrix with $\mu_{ij}-\b_j^\trans\h_i$ as the $(i,j)$-th entry, and $\bSigma_i=\mathrm{diag}(\sigma_{i1}^2,\cdots, \sigma_{ip}^2)$.} 
In the following, we develop joint optimization and separate optimization to update $(\bb, \boldsymbol\varsigma)$, in which both methods produce iterative explicit forms for $(\bb, \boldsymbol\varsigma)$.

{\noindent\bf\underline{\noindent Joint optimization method}}. Define {\red
\begin{equation}\label{lse_beta}
\W(\bb)=\frac{1}{n}\left\{(\widetilde\Y^{\trans}- \bb\Z^{\trans})^{\otimes 2}+ \sum_i \bSigma_i\right\},
\end{equation}
where $A^{\otimes 2}=AA^{\trans} $ for any matrix $A$. }
Then, the optimization problem \eqref{eq:sumBeta} can be reformulated as
\vspace{-0.2in}
\begin{equation}\label{eq:matBeta}
  \max_{\bb, \boldsymbol\varsigma} \frac{n}{2}\left[\ln |\bSigma_{\varepsilon\varepsilon}^{-1}(\boldsymbol\varsigma)|-\mathrm{Tr}\{\bSigma_{\varepsilon\varepsilon}^{-1}(\boldsymbol\varsigma)\W(\bb)\}\right]  \mbox{ subject to rank}(\bb)=r,
\end{equation}
where ``$\mathrm{Tr}$" is the trace operator for a matrix.
The following theorem, based on matrix analysis theory, ensures the explicit global maximizer.
\nvs
\begin{Theorem}\label{th:joint}
The optimization problem \eqref{eq:matBeta} has a global maximizer $(\bb_{*},{\boldsymbol\varsigma}_{*})$ given by
\nvs
\begin{eqnarray}
     && \bb_{*} =  \tilde\bSigma_{\varepsilon\varepsilon}^{1/2} \V_{(r)}^{\otimes 2}\tilde\bSigma_{\varepsilon\varepsilon}^{-1/2} \tilde\bb, \label{eq:betajoint} \\
    && {\boldsymbol\varsigma}_{*} = \frac{1}{n}\mathrm{vecdiag}\{(\widetilde\Y^{\trans}- \bb_{*}\Z^{\trans})^{\otimes 2}+  \sum_i \bSigma_i\}, \label{eq:varsigmajoint}
\end{eqnarray}
where $\tilde\bSigma_{\varepsilon\varepsilon}=\frac{1}{n}\{ (\widetilde\Y^{\trans}- \tilde\bb\Z^{\trans})^{\otimes 2} +  \sum_i \bSigma_i\}$, {\red $\tilde\bb=\widetilde\Y^{\trans}\Z(\Z^{\trans}\Z)^{-1}$}, $\V_{(r)}$ is the matrix consisting of the eigenvectors of $\tilde\bSigma_{\varepsilon\varepsilon}^{-1/2}\tilde\bb(n^{-1}\Z^{\trans}\Z)\tilde\bb^{\trans}\tilde\bSigma_{\varepsilon\varepsilon}^{-1/2}$ that correspond to the first $r$ largest eigenvalues, and $\mathrm{vecdiag}$ is an operator that extract the diagonal elements of a square matrix and stack them into a column vector.
\end{Theorem}

{\bf\noindent\underline{\noindent Separate optimization method}}. We first consider the low-rank estimator for $\bb$ given ${\boldsymbol\varsigma}$ by solving the optimization problem derived from \eqref{eq:sumBeta},
\vspace{-0.2in}
\begin{equation}\label{eq:matBeta2}
\max_{\bb} -\mathrm{Tr}\left\{\bSigma^{-1}_{\varepsilon\varepsilon}(\boldsymbol\varsigma)\frac{1}{n}(\widetilde\Y^{\trans}-\bb\Z^{\trans})^{\otimes 2}\right\}, \mbox{ subject to rank}(\bb)=r.
\end{equation}
\vspace{-0.6in}
\begin{Pro}\it
The optimization problem \eqref{eq:matBeta2} has a global maximizer $\bb^{*}$ given by
\vspace{-0.2in}
\begin{eqnarray} \label{eq:betasep}
     && \bb^{*} =  \bSigma^{1/2}_{\varepsilon\varepsilon}(\boldsymbol\varsigma) \bar  \V_{(r)}\bar  \V_{(r)}^{\trans}\bSigma^{-1/2}_{\varepsilon\varepsilon}(\boldsymbol\varsigma) \tilde\bb,
\end{eqnarray}
where $\tilde\bb$ is defined in Theorem \ref{th:joint}, and $\bar  \V_{(r)}$ is the matrix consisting of eigenvectors of \\ $\bSigma_{\varepsilon\varepsilon}^{-1/2}(\boldsymbol\varsigma)\tilde\bb(n^{-1}\Z^{\trans}\Z)\tilde\bb^{\trans}\bSigma_{\varepsilon\varepsilon}(\boldsymbol\varsigma)^{-1/2}$ that correspond to its first $r$ largest eigenvalues.
\end{Pro}
\vspace{-0.1in}
Then, plugging the estimator $\bb^{*}$ of $\bb$ into \eqref{eq:sumBeta} yields the update of $\boldsymbol\varsigma$
\nvs
\begin{eqnarray} \label{eq:varsigmasep}
     && {\boldsymbol\varsigma}^{*}= \mathrm{diagvec}\{\frac{1}{n} (\widetilde\Y^{\trans}-\bb^*\Z ^{\trans})^{\otimes 2}+  \sum_i \bSigma_i\}.
\end{eqnarray}
If dividing the joint optimization of $(\bb, \boldsymbol\varsigma)$ into two coordinate blocks, then the separate optimization method in \eqref{eq:betasep} and \eqref{eq:varsigmasep} corresponds to the coordinate ascent form of joint optimization in \eqref{eq:betajoint} and \eqref{eq:varsigmajoint}. Let $\bar  \A=\bSigma_{\varepsilon\varepsilon}(\boldsymbol\varsigma)^{-1/2}\tilde\bb(n^{-1}\Z^{\trans}\Z)^{1/2}\in \mathbb{R}^{p\times d}$. Both of these optimization methods involve computing the eigenvectors of $\bar   \A \bar   \A^{\trans}$, resulting in the computational complexity of {$O(p^3)$}. However, the joint optimization method requires an additional computation of the inverse of a dense $p\times p$ matrix, i.e., $\tilde\bSigma_{\varepsilon\varepsilon}^{-1/2}$. Therefore, we opt to implement the separate optimization method in our algorithm.  To further reduce the computational burden, we estimate the eigenvectors of $\bar  \A \bar  \A^{\trans}$ through a singular value decomposition (SVD) on $\bar  \A$ instead of an eigendecomposition on $\bar  \A \bar  \A^{\trans}$. This procedure reduces the computation complexity to $O(pd^2)$ if $p\geq d$. Furthermore, we utilize the approximate SVD method~\citep{lewis2021irlba}  to expedite the computation of the SVD on $\bar  \A$.

This proposed variational EM algorithm is highly practical, as each parameter has an explicit iterative solution that is easy to implement. The algorithm is summarized in Algorithm \ref{alg:icmem}. {\red To exert the identifiability conditions on $\H^{(t)}$ and $\B^{(t)}$, we first obtain the residual matrix, $\E$, by conducting multi-response linear regression using $\H^{(t)}$ as response matrix and $\Z$ as covariate matrix, i.e., $\H^{(t)}=\Z \wh\balpha + \E$. Then we perform singular value decomposition on $\E \B^{(t),\trans} = \U \V \D^{\trans}$, while ensuring that the first nonzero element of each column of $\D$ is positive. Subsequently, let $\H^{(t)}= \sqrt{n} \U$ and $\B^{(t)}= \D \V /\sqrt{n}$}.
Let $\mathcal{G}$ be the parameter space of the parameter satisfying Conditions (A1)--(A3). The following theorem demonstrates that the proposed iterative algorithm converges.
\begin{Theorem}
If conditions (S1)--(S2) in the Supporting Information hold,  given the proposed variational EM algorithm, we have that
all the limit points of $(\btheta^{(t)}, \bg^{(t)})$ are local maxima of $ELBO(\btheta, \bg)$ in the parameter space $\mathcal{G}$, and $ELBO(\btheta, \bg)$ converges monotonically to $L^{*}=ELBO(\btheta^{*}, \bg^{*})$ for some $(\btheta^{*}, \bg^{*})\in\mathcal{G}_{*}$, where $\mathcal{G}_{*}=\{\mbox{set of local maxima in the interior}$ $\mbox{ of }~\mathcal{G}\}.$
\end{Theorem}
\nvs
\renewcommand{\algorithmicrequire}{\textbf{Input:}}
	\renewcommand{\algorithmicensure}{\textbf{Output:}}
\begin{algorithm}
	\small
	\caption{The proposed variational EM algorithm for COAP}
	\label{alg:icmem}
	\begin{algorithmic}[1]
		\Require$\X$, $\Z$, $q$, $r$, maximum iterations $maxIter$, relative tolerance of ELBO ($epsELBO$).
		\Ensure  {\red $\wh\bb, \wh\H, \wh\B, \wh{\boldsymbol\varsigma}$}
		\State Initialize {\red $\bg^{(0)} = (\mu^{(0)}_{ij},\sigma^{2,(0)}_{ij},i=1,\cdots, n, j=1,\cdots, p)$ and  $\btheta^{(0)}=(\bb^{(0)}, \B^{(0)},\H^{(0)},\boldsymbol\varsigma^{(0)}) $.}
		\For{ each $t = 1, \cdots, maxIter $}
		\State Update variational parameters $\bg^{(t)}$ {\red based on Equation~\eqref{eq:updateY}};
		\State Update model parameters $\btheta^{(t)}$ based on {\red Equations~\eqref{eq:updata_bj}, \eqref{eq:updata_hi}, \eqref{eq:betasep} and \eqref{eq:varsigmasep}};
        \State Evaluate the evidence lower bound $ELBO_t=ELBO(\btheta^{(t)}, \bg^{(t)})$ by \eqref{eq:elbo}.
        \If {$|ELBO_t - ELBO_{t-1}|/ |ELBO_{t-1}| < epsELBO$}
       \State  break;
        \EndIf
		\EndFor
		\State  Exert the identifiability conditions (A1)--(A3) on $\H^{(t)}$ and $\B^{(t)}$.
		\State \textbf{return} $\wh\bb=\bb^{(t)}, \wh\H=\H^{(t)}, \wh\B=\H^{(t)}, \wh{\boldsymbol\varsigma}=\boldsymbol\varsigma^{(t)}$.
	\end{algorithmic}
\end{algorithm}


\subsection{Tuning parameter selection}\label{selectpara}
The number of factors ($q$) and the rank of $\bb$ ($r$) are both undetermined tuning parameters that require selection. Different from existing methods of selecting the number of factors in the linear factor models, we propose to account for the nonlinear factor structure via a surrogate covariance matrix. In particular, we extend the eigenvalue ratio (EVR) based method \citep{ahn2013eigenvalue}, and propose a singular value ratio (SVR) method for determining both $q$ and $r$, which is simple to implement.

The EVR method estimates $q$ by $\widehat q = \arg\max_{k\leq q_{\max}} \frac{\lambda_k(\widehat\bSigma_x)}{\lambda_{k+1}(\widehat\bSigma_x)}$, where $\widehat\bSigma_x$ is the sample covariance of $\x_i$ in the linear factor model setting, $\lambda_k(\widehat\bSigma_x)$ is the $k$-th largest eigenvalue of $\widehat\bSigma_x$, and  $q_{\max}$ is the upper bound for $q$. 
Due to the absence of a linear factor structure, we employ a surrogate of $\widehat\bSigma_x$, denoted as $\widehat\bSigma_{hb}$, which is defined as the sample covariance matrix of $\wh\B\wh\h_i$.  By the identifiable conditions (A1)--(A3), we have $\widehat\bSigma_{hb}=\wh\B\wh\B^{\trans}$. The SVR estimator of $q$ is obtained as follows. {\red  We first fit our model with $q=q_{\max}$ and $r=r_{\max}$, then  estimate $q$ by
$\widehat q= \arg\max_{k\leq q_{\max}} \frac{\nu_k(\wh\B)}{\nu_{k+1}(\wh\B)}$}, where $r_{\max}$ is the upper bound for $r$, and $\nu_k(\wh\B)$ is the $k$-largest singular value of $\wh\B$. We refer to this as the singular value ratio based method.
Following the same principle, we estimate the value of $r$ by calculating the singular value ratio of $\wh\bb$. 


\nvs
\section{Simulation study}\label{sec:simu}
In this section, we demonstrate the effectiveness of our proposed COAP through simulation studies by conducting $N=200$ realizations. We compare COAP with various prominent methods in the literature. They are

~(1) High-dimensional LFM~\citep{bai2002determining} implemented in the R package {\bf GFM};

~(2) PoissonPCA~\citep{kenney2021poisson} implemented in the R package {\bf PoissonPCA};

~(3) Zero-inflated Poisson factor model~\citep[ZIPFA,][]{xu2021zero} implemented in the R package {\bf ZIPFA};

~(4) Generalized factor model~\citep{liu2023generalized} implemented in the R package {\bf GFM};

~(5) PLNPCA~\citep{chiquet2018variational} implemented in the R package {\bf PLNmodels};

{\red ~(6) Generalized linear
latent variable models~\citep[GLLVM, ][]{hui2017variational} implemented in the R package {\bf gllvm}.}

~(7) Poisson regression model for each $x_{ij}(j=1,\cdots,p)$, implemented in {\bf stats} R package;

~(8) Multi-response reduced-rank Poisson regression model ~\citep[MMMR,][]{luo2018leveraging} implemented in {\bf rrpack} R package.

{\red Details of the compared can be referred to Appendix D.1 in Supporting Information. In the implementation of the compared methods, we maintain the default settings of the R packages and solely adjust the number of factors/PCs.} We evaluate the accuracy of loading matrix $\B$ and factor matrix $\H$ in terms of the trace statistic~\citep{doz2012quasi}, i.e., $\mathrm{Tr}(\wh\H,\H_0)=\frac{\mathrm{Tr}\{\H_0^{\trans} \wh\H(\wh\H^{\trans}\wh\H)^{-1} \H_0^{\trans}\wh\H\}}{\mathrm{Tr}(\H_0^{\trans}\H_0)}$, and assess the estimation accuracy (EA) of $\bb_0$ and $\bb_{.1}$ by employing $\mathrm{EA}_{\bb} = \frac{1}{N} \sum_i \|\wh\bb^{(i)} - \bb_0\|$ and $\mathrm{EA}_{\bb_{.1}} = \frac{1}{N} \sum_i \|\wh\bb_{.1}^{(i)} - \bb_{.1,0}\|$, where $\bb_{.1}$ is the first column of $\bb$. $\mathrm{EA}_{\bb_{.1}}$ is calculated since methods (1)--(4) are only able to estimate the intercepts. The trace statistic is a metric that ranges from 0 to 1, and it is considered better when it has a larger value. On the other hand, the EA metric takes non-negative values and is considered better when it has a smaller value.
\nvs
\subsection{Data generation}
We simulate data from models \eqref{eq:xij} and \eqref{eq:yij}, i.e., $x_{ij}|\widetilde   y_{ij} \sim  Pois(a_i \widetilde   y_{ij}), \ln (\widetilde   y_{ij}) = \z_i^\trans\bb_j + \h_i^\trans\b_j + \varepsilon_{ij}$. We set $a_i=1,\forall i$ by default, since we focus on the absolute level of $x_{ij}$. $\breve{\z}_i$ is independently generated from $N(\0_{d-1},(0.5^{|i-j|})_{(d-1)\times (d-1)}), \z_i = (1, \breve{\z}_i^\trans)^\trans$, $\bb_0=4\rho_{z}\U_1 \V_1^{\trans}/p $ with $\U_1\in \mathbb{R}^{d\times r_0}$ and $\V_1 \in \mathbb{R}^{p\times r_0}$, where each element of $\U_1$ and $\V_1$ is drawn from standard Gaussian distribution, and $\rho_z$ is a scalar to control the signal strength. Next, we generate $\breve{\h}_i$ from $N(\0_q,(0.5^{|i-j|})_{q\times q})$ and denote $\breve{\H}=(\breve{\h}_1,\cdots,\breve{\h}_n)^{\trans}$, regress  $\breve{\H}$ on $\Z$ to obtain the residual $\E$, and perform column orthogonality for $\E$ to obtain $\H_0$ such that $\H_0 \perp \Z$. To generate $\B_0$, we first generate $\breve{\B}=(\breve{b}_{jk})\in \mathbb{R}^{p\times q}$ with $\breve{b}_{jk}\stackrel{i.i.d.}\sim N(0,1)$, then perform SVD $\breve{\B}=\U_2 \bLambda_2 \V_2^{\trans}$, and let $\B_0= \rho_\B \U_2 \bLambda_2/ \rho_{\max}$, where $\rho_\B$ is a scalar to control the signal strength and $\rho_{\max}$ is the maximum element of $ \U_2 \bLambda_2$. Note that $\H_0$ and $\B_0$ satisfy the identifiable conditions (A1)--(A3) given in Section \ref{sec:model}. Finally, we set $\varepsilon_{ij} \stackrel{i.i.d.}\sim N(0, \sigma^2)$.
After generating $\bb_0$ and $\B_0$, they are fixed in  repetition. We set $d=50,r_0=6, q_0=5$ and $\sigma^2=1$ without specified.
\nvs
\subsection{Simulation results}
{\bf \underline{Scenario 1}}. To investigate the performance of the estimated regression parts and factor parts as $n$ or $p$ increases, we generate data with fixed $p=200$ and varied $n\in \{100, 250, 400\}$, or fixed $n=200$ and varied $p\in \{100, 250, 400\}$. We set $(\rho_z, \rho_{\B})=(6,3)$ or $(10, 2)$ to assess the impact of signal strength of $\Z\bb^{\trans}$ or $\H\B^{\trans}$ in the matrix form of model \eqref{eq:yij}, i.e., $\Y=\Z\bb^{\trans} + \H\B^{\trans} + \E$. {\red We refer to this particular data setting as Scenario 1.1.}

The standard MRRR is written as $g(\mathrm{E}(\X|\Z))=\boldsymbol\Theta$ with $\boldsymbol\Theta= \Z\bb^{\trans}$, which ignores the factor structure $\H\B^{\trans}$, denoted by MRRR$_Z$, where $g$ is the log link. We consider a variant of MMMR$_Z$ by setting $\boldsymbol\Theta= \Z\bb^{\trans} + \widetilde   \X \boldsymbol\Gamma$, with $\tilde\X = \I_n$ and $\boldsymbol\Gamma=\H\B^{\trans}$, denoted by MMMR$_F$. We investigate the average (standard deviation) of $\mathrm{EA}_{\bb_{.1}},\mathrm{EA}_{\bb},\mathrm{Tr}(\wh\H,\H_0)$ and $\mathrm{Tr}(\wh\B,\B_0)$ for the COAP and  {\red nine} comparison methods that include two versions of MRRR. Tables \ref{Scen1_n} and \ref{Scen1_p} present the estimated results based on 200 repetitions with the true number of factors and true rank of $\bb$ for all methods. {\red Notably, when $p=400$, GLLVM fails to converge, resulting in unavailable related results}. From Tables \ref{Scen1_n} and \ref{Scen1_p},  we can see that, the average $\mathrm{EA}_{\bb_{.1}}$ and $\mathrm{EA}_{\bb}$ of COAP are consistently smaller, and $\mathrm{Tr}(\wh\H,\H_0)$ and $\mathrm{Tr}(\wh\B,\B_0)$ of COAP are consistently larger than those of the nine compared methods over all considered cases, which suggests the COAP is superior to existing methods. It is unsurprising since each of the compared methods has its own set of limitations. For instance, LFM fails to account for both count variables and covariates, while  PoissonPCA, GFM and ZIPFA are unable to take into account covariates. GLM and MRRR$_Z$ are not equipped to handle factor structures, while PLNPCA {\red and GLLVM} fall short in accounting for the low-rank structure of the coefficient matrix. Lastly,  MRRR$_F$ treats the coefficient matrix and factor part as a single entity, and estimates a low-rank matrix of size $p \times (d+n)$ in a $(r+q)(p+d+n)$ dimensional space, whereas COAP estimates the two low-rank matrices separately in a much smaller $(p+d)r+ (n+p)q$ dimensional space.

Compared to the results obtained with $(\rho_z, \rho_{\B})=(6,3)$, the estimation accuracy of $\bb$ is improved for $(\rho_z, \rho_{\B})=(10,2)$, while the precision of $\wh\H$ and $\wh\B$ decrease. Notably, the improvement of COAP over GFM is more significant. These findings align with the underlying signal balance between $\Z\bb^{\trans}$ and $\H\B^{\trans}$. Specifically, the signal of the regression term $\Z\bb^{\trans}$ increases, while the signal of the factor term $\H\B^{\trans}$ decreases for $(\rho_z, \rho_{\B})=(10,2)$ when compared to $(\rho_z, \rho_{\B})=(6,3)$.
Tables \ref{Scen1_n} and \ref{Scen1_p} demonstrate that the precision of COAP for estimating $\wh\bb$ and $\wh\B$ improves with increasing $n$ and is less sensitive to $p$. Meanwhile, the precision of $\wh\H$ increases with increasing $p$ and is less sensitive to $n$. This is because the estimators $\wh\bb_j$ and $\wh\b_j$ leverage the information of the $j$th variable of $n$ individuals, while the estimation of $\h_i$ is based on the information of $p$ variables of the $i$th individual.
\nvs
\begin{table}
 \centering \renewcommand\tabcolsep{4pt}
  \caption{ \red Comparison of COAP and other methods for parameter estimation.  Reported are the average (standard deviation) for performance metrics in Scenario 1.1 with fixed $p$ and varied $n$.}
   \small
    \begin{tabular}{rlrrrrrr}
    \hline
    &&\multicolumn{3}{c}{$(\rho_z, \rho_{\B})=(6,3)$} &\multicolumn{3}{c}{$(\rho_z, \rho_{\B})=(10,2)$}\\
    \cmidrule(lr){3-5}\cmidrule(lr){6-8}
              &(n,p)         & \multicolumn{1}{l}{(100,200)}        & \multicolumn{1}{l}{(250,200)}       & \multicolumn{1}{l}{(400,200)} & \multicolumn{1}{l}{(100,200)}        & \multicolumn{1}{l}{(250,200)}       & \multicolumn{1}{l}{(400,200)}  \\
        \cmidrule(lr){2-5}\cmidrule(lr){6-8}
   COAP & $\mathrm{EA}_{\bb_{.1}}$ &0.41(0.01)   &0.40(6e-3)  &0.40(0.01)     &0.29(0.01)   &0.28(0.01)   &0.28(4e-3) \\
             & $\mathrm{EA}_{\bb}$ &0.11(5e-3) &0.07(1e-3) &0.06(7e-4) &0.09(5e-3)&0.05(1e-3) &0.05(5e-4)\\
             & $\mathrm{Tr}(\wh\H,\H_0)$ &0.97(3e-3) &0.95(2e-3)   &0.95(2e-3)  &0.92(0.02)   &0.89(4e-3) &0.89(4e-3)\\
             & $\mathrm{Tr}(\wh\B,\B_0)$ &0.85(0.01)   &0.93(4e-3)   &0.95(2e-3) &0.73(0.03)   &0.88(0.01)   &0.92(4e-3) \\
          \cmidrule(lr){2-5}\cmidrule(lr){6-8}
    LFM & $\mathrm{EA}_{\bb_{.1}}$ &14.2(1.68)   &14.0(0.71)   &14.0(0.60)     &4.67(0.20)   &4.63(0.11)   &4.62(0.08) \\
             & $\mathrm{Tr}(\wh\H,\H_0)$ &0.28(0.04)   &0.23(0.04)   & 0.22(0.04)       & 0.23(0.04)   & 0.18(0.04)   & 0.18(0.03) \\
             & $\mathrm{Tr}(\wh\B,\B_0)$
 &0.10(0.02)   &0.09(0.02)   &0.09(0.02)    &0.11(0.02)   &0.11(0.02)   &0.11(0.02) \\
\cmidrule(lr){2-5}\cmidrule(lr){6-8}
    PoissonPCA &$\mathrm{EA}_{\bb_{.1}}$ & 1.53(0.01)   & 1.53(0.01)   & 1.53(0.01)    & 1.26(0.01)   & 1.27(0.01)   & 1.27(5e-3)
\\
             & $\mathrm{Tr}(\wh\H,\H_0)$ &0.25(0.04)   &0.20(0.04)   &0.19(0.03)      &0.21(0.04)   &0.16(0.03)   &0.16(0.03) \\
             & $\mathrm{Tr}(\wh\B,\B_0)$
 &0.10(0.02)   &0.09(0.02)   &0.09(0.02)         &0.11(0.03)   &0.11(0.02)   &0.11(0.02) \\
\cmidrule(lr){2-5}\cmidrule(lr){6-8}
    PLNPCA & $\mathrm{EA}_{\bb_{.1}}$& 2.60(0.82)   & 1.44(0.23)   & 0.95(0.23)     & 1.77(0.10)   & 0.39(0.13)   & 0.33(0.01) \\
             & $\mathrm{EA}_{\bb}$ & 0.77(0.08)   & 0.35(0.04)   &0.22(0.04)     &0.65(0.02)   &0.17(0.02)   &0.13(2e-3) \\
             & $\mathrm{Tr}(\wh\H,\H_0)$ &0.88(0.01)   &0.83(0.01)   &0.82(0.01)          &0.86(0.01)   &0.79(0.01)   &0.77(0.01) \\
             & $\mathrm{Tr}(\wh\B,\B_0)$ &0.63(0.02)   & 0.81(0.03)   &0.90(0.02)       &0.54(0.02)   & 0.82(0.02)   &0.88(0.01) \\
   \cmidrule(lr){2-5}\cmidrule(lr){6-8}
    GFM & $\mathrm{EA}_{\bb_{.1}}$& 0.68(0.01)   &0.66(0.01)   & 0.66(0.01)   &0.41(0.01)   &0.39(0.01)   &0.39(0.01) \\
             &$\mathrm{Tr}(\wh\H,\H_0)$ & 0.93(4e-3) & 0.94(3e-3)& 0.94(2e-3) & 0.77(0.04)   &0.82(0.02)   &0.83(0.02) \\
             & $\mathrm{Tr}(\wh\B,\B_0)$ & 0.83(0.01)   & 0.90(4e-3) & 0.92(4e-3) &0.68(0.04)   &0.83(0.02)   &0.87(0.01) \\
    \cmidrule(lr){2-5}\cmidrule(lr){6-8}
    ZIPFA & $\mathrm{Tr}(\wh\H,\H_0)$ & 0.60(0.06)   & 0.61(0.02)   & 0.61(0.02)    & 0.41(0.04)   & 0.45(0.03)   & 0.46(0.02) \\
             & $\mathrm{Tr}(\wh\B,\B_0)$ & 0.53(0.03)   & 0.63(0.02)   & 0.66(0.01)   & 0.36(0.03)   & 0.51(0.03)   & 0.57(0.02) \\
    \cmidrule(lr){2-5}\cmidrule(lr){6-8}
    MRRR$_F$ & $\mathrm{EA}_{\bb_{.1}}$& 1.40(1.23)   & 1.25(0.22)   & 1.19(0.13)    & 0.67(0.06)   & 0.68(0.03)   & 0.71(0.02) \\
             &$\mathrm{EA}_{\bb}$& 0.21(0.17)   & 0.19(0.03)   & 0.18(0.02)   & 0.12(5e-3) & 0.11(3e-3) & 0.11(3e-3) \\
             &$\mathrm{Tr}(\wh\H,\H_0)$& 0.20(0.04)   & 0.17(0.04)   & 0.16(0.03)     & 0.20(0.04)   & 0.16(0.03)   & 0.14(0.03) \\
             &$\mathrm{Tr}(\wh\B,\B_0)$ & 0.04(0.01)   & 0.05(0.01)   & 0.05(0.01)       & 0.03(0.01)   & 0.04(0.01)   & 0.04(0.01) \\
   \cmidrule(lr){2-5}\cmidrule(lr){6-8}
    GLM &$\mathrm{EA}_{\bb_{.1}}$& 113.8(206) & 37.20(125) & 17.92(90.0)   & 83.97(147) & 1.11(4.28)   & 0.50(0.01) \\
             &$\mathrm{EA}_{\bb}$ & 44.7(82.5)  & 7.89(25.8)  & 3.53(18.3)   & 31.8(54.0)  & 0.32(0.92)   & 0.14(2e-3) \\
    \cmidrule(lr){2-5}\cmidrule(lr){6-8}
   MRRR$_Z$ & $\mathrm{EA}_{\bb_{.1}}$ & 1.09(0.01)   & 1.10(0.01)   & 1.10(0.01)     & 0.76(0.01)   & 0.77(0.01)   & 0.77(0.01) \\
             & $\mathrm{EA}_{\bb}$& 0.16(2e-3)& 0.16(1e-3)& 0.16(1e-3) & 0.12(1e-3)& 0.12(1e-3) & 0.12(7e-4) \\
                 \cmidrule(lr){2-5}\cmidrule(lr){6-8}
   GLLVM & $\mathrm{EA}_{\bb_{.1}}$ & 9.04(2.50)   & 3.16(1.26)  & 1.63(0.81)     & 7.34(2.67)  & 0.49(0.38) & 0.33(0.01) \\
             & $\mathrm{EA}_{\bb}$ & 2.91(0.77)& 0.69(0.27) & 0.34(0.14) & 2.50(0.93)& 0.19(0.06) & 0.12(2e-3)\\
             & $\mathrm{Tr}(\wh\H,\H_0)$ & 0.91(0.04) & 0.85(0.02)   & 0.84(0.02)  & 0.86(0.02)   & 0.80(0.01) & 0.78(0.01) \\
             & $\mathrm{Tr}(\wh\B,\B_0)$ & 0.52(0.04)  & 0.77(0.06)   & 0.89(0.04) & 0.47(0.03)  & 0.82(0.02) & 0.88(0.01) \\
       \hline
        \end{tabular}
         \label{Scen1_n}%
\end{table}
\nvs
\begin{table}
 \centering \renewcommand\tabcolsep{4pt}
  \caption{\red Comparison of COAP and other methods for parameter estimation.  Reported are the average (standard deviation) for performance metrics in Scenario 1.1 with fixed $n$ and varied $p$. ``$-$" means that the algorithm of the corresponding method breaks down under the corresponding data setting.}
   \small
    \begin{tabular}{rlrrrrrr}
    \hline
    &&\multicolumn{3}{c}{$(\rho_z, \rho_{\B})=(6,3)$} &\multicolumn{3}{c}{$(\rho_z, \rho_{\B})=(10,2)$}\\
    \cmidrule(lr){3-5}\cmidrule(lr){6-8}
              &(n,p)         & \multicolumn{1}{l}{(200,100)}        & \multicolumn{1}{l}{(200,250)}       & \multicolumn{1}{l}{(200,400)} & \multicolumn{1}{l}{(200,100)}        & \multicolumn{1}{l}{(200,250)}       & \multicolumn{1}{l}{(200,400)}  \\
        \cmidrule(lr){2-5}\cmidrule(lr){6-8}
  COAP & $\mathrm{EA}_{\bb_{.1}}$ & 0.34(0.01)   & 0.49(0.01)   & 0.45(0.01)   & 0.27(0.01)  & 0.31(0.01)   & 0.30(0.01)  \\
             & $\mathrm{EA}_{\bb}$ & 0.07(2e-3) & 0.09(2e-3) & 0.08(1e-3) & 0.06(2e-3) & 0.06(1e-3) & 0.06(2e-3) \\
             &$\mathrm{Tr}(\wh\H,\H_0)$& 0.95(4e-3) & 0.98(1e-3) & 0.98(8e-4) & 0.84(0.04)  & 0.95(3e-3)  & 0.97(2e-3)  \\
             &$\mathrm{Tr}(\wh\B,\B_0)$& 0.96(3e-3)& 0.93(3e-3)& 0.93(2e-3) & 0.83(0.05)  & 0.89(5e-3) & 0.90(3e-3)\\
    \cmidrule(lr){2-5}\cmidrule(lr){6-8}
    LFM &$\mathrm{EA}_{\bb_{.1}}$
    & 79.2(15.2)  & 27.8(4.40)   & 17.9(1.03)   & 12.84(1.71)   & 6.08(0.19)   & 4.81(0.10)  \\
&$\mathrm{Tr}(\wh\H,\H_0)$& 0.19(0.03)   & 0.21(0.04)   & 0.25(0.05)   & 0.14(0.03)  & 0.25(0.04)  & 0.31(0.06)   \\
   &$\mathrm{Tr}(\wh\B,\B_0)$   & 0.11(0.01)   & 0.09(0.01)   & 0.08(0.02)   & 0.11(0.02)   & 0.12(0.02)   & 0.13(0.03)  \\
    \cmidrule(lr){2-5}\cmidrule(lr){6-8}
   PoissonPCA &$\mathrm{EA}_{\bb_{.1}}$
    & 1.80(0.02)   & 1.75(0.01)   & 1.72(0.01)   & 1.52(0.02)   & 1.36(0.01)   &1.33(4e-3)  \\
     &$\mathrm{Tr}(\wh\H,\H_0)$ & 0.18(0.03)   & 0.20(0.03)   & 0.23(0.04)   & 0.13(0.03)   & 0.23(0.04)   & 0.28(0.06)  \\
    &$\mathrm{Tr}(\wh\B,\B_0)$ & 0.11(0.01)   & 0.09(0.01)   & 0.08(0.02)   & 0.11(0.02)   & 0.12(0.02)   & 0.14(0.03)  \\
    \cmidrule(lr){2-5}\cmidrule(lr){6-8}
    PLNPCA &$\mathrm{EA}_{\bb_{.1}}$
    & 0.68(0.20)   & 1.73(0.14)   & 1.51(0.47)   & 0.26(0.05)   & 0.85(0.17)   & 0.67(0.10)  \\
        &$\mathrm{EA}_{\bb}$     & 0.27(0.02)   & 0.42(0.03)   & 0.38(0.05)   & 0.20(0.01)   & 0.28(0.03)   & 0.24(0.01)  \\
      &$\mathrm{Tr}(\wh\H,\H_0)$       & 0.79(0.02)   & 0.87(0.01)   & 0.92(0.01)   & 0.76(0.01)   & 0.87(0.01)  & 0.92(5e-3)  \\
       &$\mathrm{Tr}(\wh\B,\B_0)$      & 0.91(0.02)   & 0.79(0.02)   & 0.83(0.02)   & 0.87(0.02)  & 0.79(0.01)   & 0.84(0.03)   \\
    \cmidrule(lr){2-5}\cmidrule(lr){6-8}
   GFM &$\mathrm{EA}_{\bb_{.1}}$
    & 0.46(0.02)   & 0.83(0.01)   & 0.72(0.01)   & 0.36(0.01)   & 0.47(0.01)   & 0.43(0.01)  \\
          &$\mathrm{Tr}(\wh\H,\H_0)$     & 0.92(5e-3) & 0.97(1e-3) & 0.98(1e-3) & 0.53(0.03) & 0.92(4e-3)   & 0.95(2e-3)  \\
       &$\mathrm{Tr}(\wh\B,\B_0)$      & 0.94(0.01)   & 0.89(5e-3) & 0.91(3e-3) & 0.63(0.03)   & 0.87(0.01)   & 0.89(4e-3)  \\
    \cmidrule(lr){2-5}\cmidrule(lr){6-8}
    ZIPFA&$\mathrm{Tr}(\wh\H,\H_0)$
     & 0.57(0.02)   & 0.67(0.01)   & 0.69(0.01)   & 0.26(0.04)  & 0.59(0.02)    & 0.66(0.01)  \\
          &$\mathrm{Tr}(\wh\B,\B_0)$   & 0.56(0.02)   & 0.61(0.02)   & 0.62(0.02)   & 0.31(0.04)   & 0.59(0.01)   & 0.61(0.01)  \\
    \cmidrule(lr){2-5}\cmidrule(lr){6-8}
   MRRR$_F$ &$\mathrm{EA}_{\bb_{.1}}$& 1.28(0.03)   & 1.53(0.42)   &2e4(2e5)  & 0.90(0.18)   & 0.79(0.06)   & 0.73(0.02)  \\
          &$\mathrm{EA}_{\bb}$     & 0.42(0.45)   & 0.23(0.07)   &2e3(3e4)  & 0.17(0.02)    & 0.13(0.01)   & 0.12(3e-3) \\
      &$\mathrm{Tr}(\wh\H,\H_0)$       & 0.16(0.03)   & 0.17(0.03)   & 0.16(0.03)   & 0.12(0.03)   & 0.22(0.04)  & 0.21(0.04)   \\
         &$\mathrm{Tr}(\wh\B,\B_0)$ & 0.09(0.01)   & 0.06(0.01)   & 0.03(0.01)   & 0.07(0.01)   & 0.04(0.01)   & 0.03(0.01)  \\
   \cmidrule(lr){2-5}\cmidrule(lr){6-8}
    GLM &$\mathrm{EA}_{\bb_{.1}}$
    & 7.56(22.2)  & 74.4(159)  & 48.5(60.5)  & 0.46(0.02)   & 64.12(537)  & 12.56(63.9)  \\
      &$\mathrm{EA}_{\bb}$     & 1.82(5.05)   & 17.5(38.5)  & 11.5(14.8)  & 0.21(111)  & 14.02(14.6)  & 2.88(4e-3)  \\
    \cmidrule(lr){2-5}\cmidrule(lr){6-8}
    MRRR$_Z$ &$\mathrm{EA}_{\bb_{.1}}$
    & 1.42(0.04)   & 1.37(0.01)   & 1.29(0.01)   & 1.08(0.03)   & 0.89(7.69)   & 0.83(0.01)  \\
          &$\mathrm{EA}_{\bb}$   & 0.21(0.01)   & 0.19(2e-3)& 0.18(2e-3) & 0.19(1.01)   &0.13(8e-4)& 0.12(3e-3)\\
                           \cmidrule(lr){2-5}\cmidrule(lr){6-8}
   GLLVM & $\mathrm{EA}_{\bb_{.1}}$ & 2.09(1.40)   & 6.99(0.48) &-(-)    & 0.26(0.06) & 2.67(1.84) &-(-) \\
             & $\mathrm{EA}_{\bb}$ & 0.55(0.28)& 1.56(0.09) &-(-)  & 0.20(0.01)& 0.67(0.41) &-(-)\\
             & $\mathrm{Tr}(\wh\H,\H_0)$ & 0.82(0.02) & 0.91(2e-3) &-(-)  & 0.77(0.01) & 0.88(0.03) &-(-)\\
             & $\mathrm{Tr}(\wh\B,\B_0)$ & 0.87(0.04) & 0.47(0.04) &-(-)  & 0.87(0.01)& 0.76(0.05) &-(-)\\
       \hline
        \end{tabular}
         \label{Scen1_p}%
\end{table}
\nvs

{\red {\bf Scenario 1.2}. Furthermore, in the Supporting Information, we examine a special case of Scenario 1.1, denoted as Scenario 1.2, where no covariate is involved. This signifies that only an intercept is present, and no covariates are generated for $\Z$. Tables S1 and S2 in the Supporting Information demonstrate COAP's superior or comparable performance compared to other methods and consistent performance patterns as in Scenario 1.1 with increasing values of $n$ or $p$. Further details are available in Appendix D.2 of Supporting Information.}

{\red {\bf Scenario 1.3}. At times, our interest lies in modeling relative counts. Therefore, we proceed to compare COAP with methods that are able to handle relative counts. We generate the data in a setup similar to Scenario 1.1, with the exception of $a_i=20$. The results presented in Table S3 of Supporting Information support the same conclusions as those obtained in Scenario 1.1. This reaffirms the outstanding performance of  COAP in both relative and absolute count Scenarios. Moreover, a comparison of Tables \ref{Scen1_n}, \ref{Scen1_p} and S3 reveals that COAP exhibits robustness to the choice of $a_i$.}

{\noindent\bf \underline{Scenario 2}}.
It is commonly observed in practice that count data exhibit overdispersion~\citep{sun2020statistical}. To investigate the performance of different estimation methods under different levels of overdispersion, we design a scenario in which we generate data from  Scenarios 1.1 and 1.2 with varying levels of dispersion parameter $\sigma^2$, i.e., $\sigma^2=1, 4$ and $8$.

We compare COAP with PLNPCA, GFM, MRRR$_F$, GLM, MRRR$_Z$ {\red and GLLVM, as these six} methods have shown better performance than other compared methods. The estimated averages (standard deviations) of $\mathrm{EA}_{\bb_{.1}},\mathrm{EA}_{\bb},\mathrm{Tr}(\wh\H,\H_0)$ and $\mathrm{Tr}(\wh\B,\B_0)$ are summarized in {\red Table \ref{tab:sce21} and Table S4 in Supporting Information}. As one can see, although the estimated accuracy of all methods decreases as $\sigma^2$ increases, the proposed COAP is more accurate than these compared methods across all settings. In contrast to the competitors, the improvement of COAP is substantial since all other methods fail to account the overdispersion caused by the extra errors.  Moreover, Scenario 2 claims similar conclusion to those from Scenario 1, that is, the precision of $\bb$ or $(\H,\B)$ increases as the signal of $\Z\bb$ or $\H\B^{\trans}$ increases, and COAP performs comparably with GFM in the setting without covaraites. We also observe that MRRR$_F$ and MRRR$_Z$ are very sensitive to the overdispersion of observed data in the sense that the estimators are invalid or the algorithm breaks down when $\sigma^2\in \{4,8\}$.
\nvs
\begin{table}
 \centering \renewcommand\tabcolsep{4pt}
  \caption{\red Comparison of COAP and other methods for parameter estimation. Reported are the average (standard deviation) of performance metrics under Scenario 2 with covariates. ``$-$" means that the algorithm of the corresponding method breaks down under the corresponding data setting.}\label{tab:sce21}
  \small
  \begin{tabular}{rlrrrrrrr}
    \hline
    &&&\multicolumn{3}{c}{$(\rho_z, \rho_{\B})=(6,3)$} &\multicolumn{3}{c}{$(\rho_z, \rho_{\B})=(10,2)$}\\
    \cmidrule(lr){3-6}\cmidrule(lr){7-9}
           &(n,p)         &   & \multicolumn{1}{l}{$\sigma^2=1$} & \multicolumn{1}{l}{$\sigma^2=4$}        & \multicolumn{1}{l}{$\sigma^2=8$}      & \multicolumn{1}{l}{$\sigma^2=1$} & \multicolumn{1}{l}{$\sigma^2=4$}        & \multicolumn{1}{l}{$\sigma^2=8$}  \\
               \cmidrule(lr){1-6}\cmidrule(lr){7-9}
   COAP &(100,200)& $\mathrm{EA}_{\bb_{.1}}$   & 0.41(0.01) & 0.59(0.02)   &0.79(0.02)& 0.29(0.01) & 0.52(0.02) & 0.72(0.02) \\
            & &$\mathrm{EA}_{\bb}$  & 0.11(5e-3) & 0.19(0.01)   &0.24(0.01) & 0.09(5e-3) & 0.17(0.01)   & 0.23(0.01) \\
            & &$\mathrm{Tr}(\wh\H,\H_0)$  & 0.97(3e-3) & 0.89(0.01)   &0.73(0.05) & 0.92(0.02)   & 0.63(0.06)   & 0.39(0.05) \\
            & & $\mathrm{Tr}(\wh\B,\B_0)$ & 0.85(0.01)   & 0.68(0.02)   &0.47(0.04) & 0.73(0.03)   & 0.37(0.04)   & 0.18(0.03) \\
     \cmidrule(lr){2-6}\cmidrule(lr){7-9}
    &(200,100) &$\mathrm{EA}_{\bb_{.1}}$& 0.34(0.01)   & 0.55(0.02)   &0.79(0.02)  & 0.27(0.01)   & 0.52(0.02)   & 0.72(0.02)  \\
            & &$\mathrm{EA}_{\bb}$  & 0.07(2e-3) & 0.13(3e-3) & 0.17(4e-3) & 0.06(2e-3) & 0.12(0.01) & 0.16(0.01) \\
             &&$\mathrm{Tr}(\wh\H,\H_0)$  & 0.95(4e-3) & 0.86(0.01)   & 0.74(0.02)  & 0.84(0.04) & 0.60(0.05)   & 0.36(0.05)  \\
            & & $\mathrm{Tr}(\wh\B,\B_0)$ & 0.96(3e-3) & 0.89(0.01)   &0.79(0.02)  & 0.83(0.05)   & 0.65(0.05)   & 0.42(0.05)  \\
 \cmidrule(lr){2-6}\cmidrule(lr){7-9}
    PLNPCA
&(100,200) &$\mathrm{EA}_{\bb_{.1}}$& 2.60(0.82)   & 1.46(0.10)   &0.91(0.07) & 1.77(0.10)   & 1.01(0.08)   & 0.74(0.05) \\
            & &$\mathrm{EA}_{\bb}$  & 0.77(0.08)   & 0.75(0.02)   &0.77(0.02)& 0.65(0.02)   & 0.69(0.02)   & 0.74(0.02) \\
           & &$\mathrm{Tr}(\wh\H,\H_0)$   & 0.88(0.01)   & 0.57(0.04)   &0.37(0.05) & 0.86(0.01)   & 0.43(0.05)   & 0.25(0.04) \\
           & & $\mathrm{Tr}(\wh\B,\B_0)$  & 0.63(0.02)   & 0.48(0.03)   & 0.30(0.04) & 0.54(0.02)   & 0.29(0.03)   & 0.15(0.02) \\
     \cmidrule(lr){2-6}\cmidrule(lr){7-9}
    &(200,100) &$\mathrm{EA}_{\bb_{.1}}$& 0.68(0.20)   & 0.70(0.05)   & 1.00(0.06)& 0.26(0.05)   & 0.63(0.03)   & 0.92(0.05)  \\
            &&$\mathrm{EA}_{\bb}$  & 0.27(0.02)   & 0.40(0.01)   &0.57(0.01)  & 0.20(0.01)   & 0.37(0.01)   & 0.54(0.01)  \\
            &&$\mathrm{Tr}(\wh\H,\H_0)$   & 0.79(0.02)   & 0.46(0.03)   & 0.28(0.03) & 0.76(0.01)   & 0.33(0.03)   & 0.17(0.03)  \\
             && $\mathrm{Tr}(\wh\B,\B_0)$ & 0.91(0.02)   & 0.67(0.04)   &0.44(0.04)  & 0.87(0.02)   & 0.46(0.04)   & 0.25(0.04)  \\
 \cmidrule(lr){2-6}\cmidrule(lr){7-9}
  GFM
  &(100,200) &$\mathrm{EA}_{\bb_{.1}}$& 0.68(0.01)   & 0.69(0.02)   & 0.87(0.02) & 0.41(0.01)   & 0.57(0.01)   & 0.77(0.02) \\
            & &$\mathrm{Tr}(\wh\H,\H_0)$ & 0.93(4e-3) & 0.79(0.02)   & 0.57(0.05) & 0.77(0.04)   & 0.44(0.05)   & 0.23(0.04) \\
              &&$\mathrm{Tr}(\wh\B,\B_0)$ & 0.83(0.01)   & 0.68(0.01)   &0.46(0.04) & 0.68(0.04)   & 0.35(0.04)   & 0.17(0.03) \\
     \cmidrule(lr){2-6}\cmidrule(lr){7-9}
    &(200,100) &$\mathrm{EA}_{\bb_{.1}}$& 0.46(0.02)   & 0.64(0.02)   &0.87(0.02)& 0.36(0.01)   & 0.58(0.01)   & 0.78(0.02)  \\
            &&$\mathrm{Tr}(\wh\H,\H_0)$  & 0.92(5e-3) & 0.82(0.01)   &0.69(0.02)  & 0.53(0.03) & 0.41(0.03)   & 0.28(0.03)  \\
            &&$\mathrm{Tr}(\wh\B,\B_0)$   & 0.94(0.01)   & 0.89(0.01)   & 0.79(0.01)  & 0.63(0.03)   & 0.52(0.04)   & 0.37(0.04)  \\
 \cmidrule(lr){2-6}\cmidrule(lr){7-9}
  MRRR$_F$
 &(100,200) &$\mathrm{EA}_{\bb_{.1}}$& 1.40(1.23)   & 1.22(0.04)   &-(-)    & 0.67(0.06)   & 0.87(0.02)   & 1.35(0.19) \\
            &&$\mathrm{EA}_{\bb}$  & 0.21(0.17)   & 0.44(0.33)   &-(-)     & 0.12(5e-3) & 0.19(0.07)   & 1.29(0.44) \\
           & &$\mathrm{Tr}(\wh\H,\H_0)$  & 0.20(0.04)   & 0.07(0.02)   &-(-)    & 0.20(0.04)   & 0.06(0.01)   & 0.05(0.01) \\
            && $\mathrm{Tr}(\wh\B,\B_0)$ & 0.04(0.01)   & 0.03(0.01)   &-(-)     & 0.03(0.01)   & 0.03(0.01)   & 0.03(0.01) \\
    \cmidrule(lr){2-6}\cmidrule(lr){7-9}
   &(200,100)&$\mathrm{EA}_{\bb_{.1}}$ & 1.28(0.03)   &-(-) &-(-)     & 0.90(0.18)   & 0.82(0.07)   & 1.67(0.28)  \\
           &&$\mathrm{EA}_{\bb}$   & 0.42(0.45)   &-(-) &-(-)    & 0.17(0.02)   & 0.37(0.25)   & 1.81(0.92)  \\
           & &$\mathrm{Tr}(\wh\H,\H_0)$  & 0.16(0.03)   &-(-)&-(-)     & 0.12(0.03)   & 0.04(0.01)   & 0.03(0.01)  \\
           && $\mathrm{Tr}(\wh\B,\B_0)$   & 0.09(0.01)   &-(-) &-(-)     & 0.07(0.01)   & 0.06(0.01)   & 0.04(0.01)  \\
 \cmidrule(lr){2-6}\cmidrule(lr){7-9}
  GLM
  &(100,200)&$\mathrm{EA}_{\bb_{.1}}$ & 113.8(206) & 76.93(128) &73.84(191) & 83.97(147) & 88.27(342) & 28.80(74.6) \\
            &&$\mathrm{EA}_{\bb}$  & 44.67(82.5)  & 30.46(63)  &26.72(61) & 31.81(54)  & 32.49(122) & 12.26(41.3) \\
    &(200,100) &$\mathrm{EA}_{\bb_{.1}}$& 7.56(22.3)  & 1.40(4.67)   & 1.33(0.06)  & 0.46(0.02)   & 0.90(0.03)   & 1.21(0.05)  \\
           & &$\mathrm{EA}_{\bb}$  & 1.82(5.05)   & 0.52(1.14)   & 0.65(0.02)  & 0.21(111)  & 0.40(0.01)   & 0.62(0.02)  \\
 \cmidrule(lr){2-6}\cmidrule(lr){7-9}
 MRRR$_Z$
  &(100,200) &$\mathrm{EA}_{\bb_{.1}}$& 1.09(0.01)   & 126(2e3) & -(-)        & 0.76(0.01)   & 1.04(0.07)   & 5e72(8e73) \\
            &&$\mathrm{EA}_{\bb}$  & 0.16(2e-3) & 17.6(233) &-(-)      & 0.12(1e-3) & 0.16(0.01)   & 7e71(1e73) \\

    &(200,100)&$\mathrm{EA}_{\bb_{.1}}$ & 1.42(0.04)   & 3e12(3e13) &-(-)       & 1.08(0.03)   & 1.50(0.15)   & 6e61(8e62) \\
            &&$\mathrm{EA}_{\bb}$   & 0.21(0.01)   & 4e11(4e12) &-(-)      & 0.19(1.01)   & 0.24(0.02)   & 8e60(1e62) \\
                    \cmidrule(lr){1-6}\cmidrule(lr){7-9}
   GLLVM &(100,200)& $\mathrm{EA}_{\bb_{.1}}$   & 9.04(2.50)& 9.25(1.37) & 10.6(2.04) & 7.34(2.67)& 6.91(1.85)& 7.01(1.66) \\
            & &$\mathrm{EA}_{\bb}$  &  2.91(0.77)& 3.40(0.46)  & 3.99(0.65)& 2.50(0.93)& 2.60(0.58)& 2.87(0.56)\\
            & &$\mathrm{Tr}(\wh\H,\H_0)$  & 0.91(0.04)& 0.59(0.07)  & 0.27(0.05)& 0.86(0.02)& 0.43(0.05)& 0.20(0.05) \\
            & & $\mathrm{Tr}(\wh\B,\B_0)$ & 0.52(0.04)& 0.25(0.04)  & 0.14(0.03)& 0.47(0.03)& 0.17(0.03)& 0.08(0.02)\\
     \cmidrule(lr){2-6}\cmidrule(lr){7-9}
    &(200,100) &$\mathrm{EA}_{\bb_{.1}}$& 2.09(1.40) & 0.83(0.80) & 1.14(0.24)& 0.26(0.06)& 0.63(0.03)& 0.99(0.19)\\
            & &$\mathrm{EA}_{\bb}$  & 0.55(0.28)& 0.44(0.15)& 0.87(0.22)& 0.20(0.01) & 0.38(0.01)& 0.74(0.12)\\
             &&$\mathrm{Tr}(\wh\H,\H_0)$  & 0.82(0.02)& 0.48(0.05)& 0.17(0.04)& 0.77(0.01)& 0.34(0.04)& 0.11(0.03)\\
            & & $\mathrm{Tr}(\wh\B,\B_0)$ & 0.87(0.04)& 0.64(0.04)& 0.37(0.04)&  0.87(0.01)& 0.44(0.04)& 0.20(0.03)\\
             \hline
        \end{tabular}
\end{table}

\begin{figure}[H]
  \centering
  \includegraphics[width=12cm]{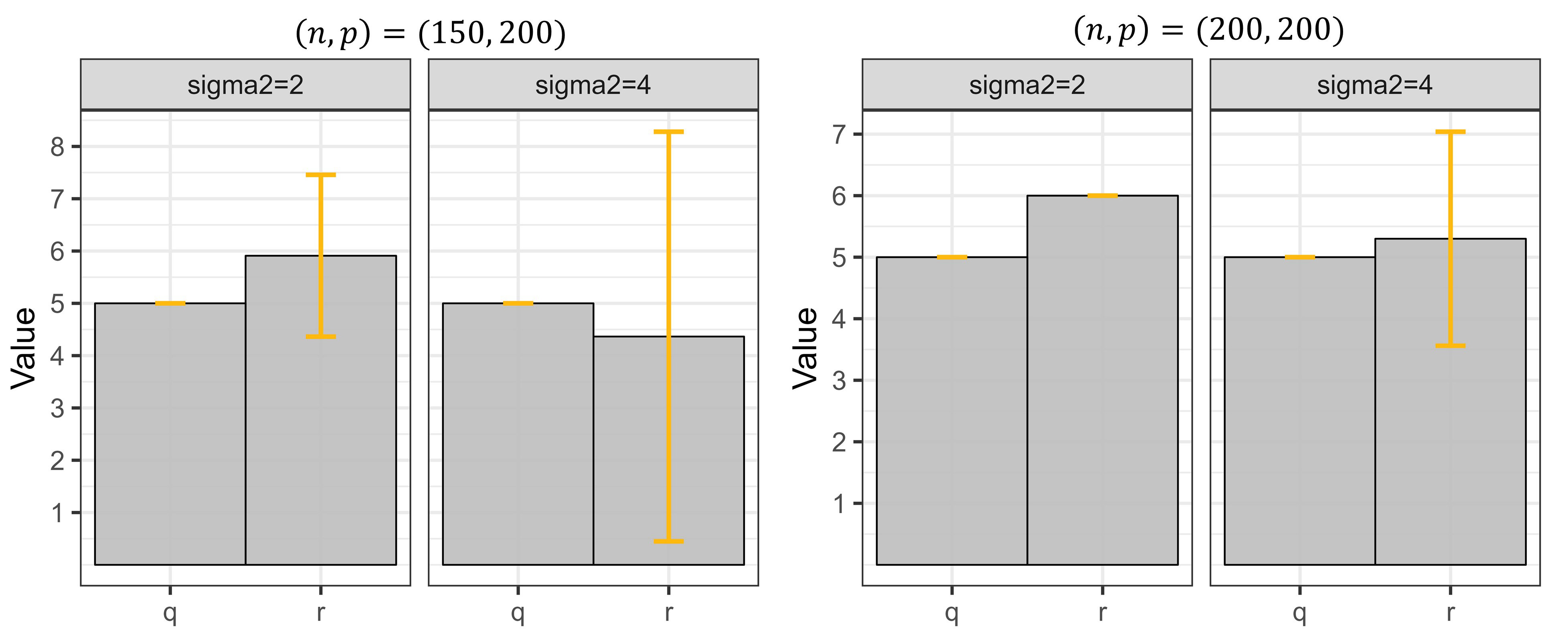}
  \caption{\red Bar plots of the average of selected $r$ and $q$ for the proposed SVR method over 200 repeats,  the error bars  represent the mean value $\pm$ standard deviation.}\label{fig:selectqrSimu}
\end{figure}

{\noindent\bf \underline{Scenario 3}}.
In this scenario, we evaluate the SVR method proposed in Section \ref{selectpara} for selecting the number of factors and the rank of $\bb$. We consider $(n,p) \in \{(150, 200), (200, 200)\}$, $\sigma^2 \in \{2, 4\}$, and keep other parameters the same as in {\red Scenario 1.1}. We set the upper bounds of $r$ and $q$ to be $r_{\max}=25$ and $q_{\max}=15$. {\red Figure \ref{fig:selectqrSimu} presents the average of selected $r$ and $q$, while Figure S2 in Supporting Information demonstrates the ratio of singular values of $\wh\bb$ and $\wh\B$ for one randomly generated data. These results} show that the SVR method can identify the underlying $q$ and $r$ with a high correction rate when the noise level is low. However, as the noise level increases, the correction rate decreases due to insufficient signal strength to accurately estimate the underlying low-rank structure. In the next scenario, we investigate the impact of incorrect selection of $q$ or $r$ on the estimation accuracy.

\nvs

{\noindent\bf \underline{Scenario 4}}.
When the signal in the analyzed data is insufficient, the estimated rank of $\bb$ or the number of factors may be incorrect. This scenario is designed to investigate the performance of the estimators when the rank of $\bb$ or the number of factors is misspecified. We follow {\red Scenario 1.1}, and  consider two cases to generate data. In Case 1, we set the true $q$ and misspecified $r=3, 9$ or $12$ in COAP and other compared methods to evaluate the influence of misselected $r$. In Case 2, we set the true $r$ and misspecified $q=3, 8$, or $11$ to evaluate the influence of misselected $q$. We compare COAP with MRRR$_F$ and MRRR$_Z$ in Case 1 since only these methods involve the selection of $r$. Similarly, we only compare COAP with LFM, PoissonPCA, PLNPCA, GFM, ZIPFA, MRRR$_F$, {\red and GLLVM} in Case 2.


The results are presented in Table \ref{Scen5_r} and {\red Table S5} in Supporting Information. By combining the results of Table \ref{Scen5_r} with those of Table \ref{Scen1_n} for $(n,p)=(100,200)$ and Table \ref{Scen1_p} for $(n,p)=(200,100)$, we observe that the estimates obtained by COAP are significantly more accurate than those obtained by MRRR$_F$ and MRRR$_Z$, regardless of whether smaller or larger rank of $\bb$ is selected. Furthermore, the estimation accuracy of COAP is insensitive to misselected $r$ when the signal of the factor term $\H\B^{\trans}$ is strong ($(\rho_z, \rho_B)=(6,3)$), while it is slightly worse when the signal of $\H\B^{\trans}$ is weak ($(\rho_z, \rho_B)=(10,2)$). A careful inspection of COAP under $(\rho_z, \rho_B)=(10,2)$ and $(n,p)=(100, 200)$ reveals that over-selection of $r$ can maintain higher performance than under-selection of $r$.

Similarly, according to {\red Table S5}, combining the results of Table \ref{Scen1_n} for $(n,p)=(100,200)$ and Table \ref{Scen1_p} for $(n,p)=(200,100)$, we observe that the estimation accuracy of COAP is consistently higher than that of LFM, PoissonPCA, PLNPCA, GFM, ZIPFA, MRRR$_F$ {and GLLVM}, regardless of selecting fewer or more factors. We also observe that the estimation accuracy of COAP with over-selected $q$ is much better than that with under-selected $q$. These findings provide a guideline for practitioners using COAP. Based on the SVR method of COAP, users can select larger values of $r$ and $q$ for more reliable and robust applications.

\nvs
\begin{table}
 \centering \renewcommand\tabcolsep{4pt}
  \caption{\red Comparison of COAP and other methods for parameter estimation when $r$ is misspecified.  Results are presented for Scenario 4 with covariates and $r_0=6$.}\label{Scen5_r}%
   \small
  \begin{tabular}{rlrrrrrrr}
    \hline
    &&&\multicolumn{3}{c}{$(\rho_z, \rho_{\B})=(6,3)$} &\multicolumn{3}{c}{$(\rho_z, \rho_{\B})=(10,2)$}\\
    \cmidrule(lr){3-6}\cmidrule(lr){7-9}
            &(n,p)         &   & \multicolumn{1}{l}{$\wh r=3$} & \multicolumn{1}{l}{$\wh r=9$}        & \multicolumn{1}{l}{$\wh r=12$}    & \multicolumn{1}{l}{$\wh r=3$} & \multicolumn{1}{l}{$\wh r=9$}        & \multicolumn{1}{l}{$\wh r=12$}  \\
               \cmidrule(lr){1-6}\cmidrule(lr){7-9}
   COAP &(100,200)& $\mathrm{EA}_{\bb_{.1}}$   & 0.41(0.01)   & 0.41(0.01)   & 0.40(0.01)   & 0.30(0.01)   & 0.28(0.01)   & 0.27(0.01)  \\
    &&$\mathrm{EA}_{\bb}$ & 0.09(4e-3) & 0.14(0.01)   & 0.15(0.01)   & 0.08(3e-3) & 0.12(0.01)   & 0.15(0.01)  \\
    &&$\mathrm{Tr}(\wh\H,\H_0)$ & 0.97(3e-3) & 0.97(3e-3) & 0.97(3e-3) & 0.91(0.03)   & 0.92(0.02)   & 0.92(0.02)  \\
    &&$\mathrm{Tr}(\wh\B,\B_0)$ & 0.85(0.01)   & 0.85(0.01)   & 0.85(0.01)   & 0.71(0.04)   & 0.73(0.03)   & 0.73(0.03)  \\
     \cmidrule(lr){2-6}\cmidrule(lr){7-9}
   &(200,100)&$\mathrm{EA}_{\bb_{.1}}$ & 0.35(0.01)   & 0.34(0.01)   & 0.34(0.01)   & 0.30(0.01)   & 0.26(0.01)   & 0.26(0.01)  \\
   &&$\mathrm{EA}_{\bb}$  & 0.08(3e-3) & 0.09(2e-3) & 0.10(2e-3) & 0.09(3e-3) & 0.08(2e-3) & 0.09(2e-3) \\
   &&$\mathrm{Tr}(\wh\H,\H_0)$  & 0.95(0.01) & 0.95(0.01) & 0.95(0.01) & 0.78(0.04)   & 0.86(0.04)   & 0.87(0.03)  \\
   &&$\mathrm{Tr}(\wh\B,\B_0)$  & 0.96(3e-3) & 0.95(3e-3) & 0.95(3e-3) & 0.71(0.04)   & 0.87(0.04)   & 0.88(0.04)  \\
    \hline
    MRRR$_F$
 &(100,200) &$\mathrm{EA}_{\bb_{.1}}$& 1.40(1.23)   & 1.40(1.25)   & 1.41(1.25)   & 0.67(0.06)   & 0.67(0.07)   & 0.67(0.07)  \\
  &&$\mathrm{EA}_{\bb}$   & 0.21(0.17)   & 0.21(0.17)   & 0.21(0.17)   & 0.12(0.01)   & 0.12(4e-3) & 0.12(0.01) \\
   &&$\mathrm{Tr}(\wh\H,\H_0)$  & 0.19(0.04)   & 0.22(0.05)   & 0.22(0.06)   & 0.18(0.04)   & 0.21(0.05)   & 0.23(0.05)  \\
    &&$\mathrm{Tr}(\wh\B,\B_0)$ & 0.04(0.01)   & 0.05(0.01)   & 0.05(0.01)   & 0.03(0.01)   & 0.04(0.01)   & 0.04(0.01)  \\
     \cmidrule(lr){2-6}\cmidrule(lr){7-9}
    &(200,100)&$\mathrm{EA}_{\bb_{.1}}$& 1.28(0.03)   & 1.28(0.03)   & 1.28(0.03)   & 0.90(0.17)   & 0.90(0.18)   & 0.90(0.18)  \\
    &&$\mathrm{EA}_{\bb}$ & 0.42(0.45)   & 0.42(0.45)   & 0.42(0.45)   & 0.17(0.02)   & 0.17(0.02)   & 0.17(0.02)  \\
    &&$\mathrm{Tr}(\wh\H,\H_0)$ & 0.16(0.03)   & 0.16(0.03)   & 0.17(0.03)   & 0.11(0.02)   & 0.12(0.03)   & 0.12(0.03)  \\
    &&$\mathrm{Tr}(\wh\B,\B_0)$ & 0.09(0.01)   & 0.09(0.01)   & 0.10(0.01)   & 0.07(0.01)   & 0.08(0.01)   & 0.08(0.01)  \\
    \hline
  MRRR$_Z$
  &(100,200) &$\mathrm{EA}_{\bb_{.1}}$  & 1.08(0.01)   & 1.08(0.01)   & 1.08(0.01)   & 0.79(0.01)   & 0.79(0.01)   & 0.79(0.01)  \\
   &&$\mathrm{EA}_{\bb}$ & 0.15(2e-3) & 0.15(2e-3) & 0.15(2e-3) & 0.12(1e-3) & 0.12(1e-3) & 0.12(1e-3) \\
     \cmidrule(lr){2-6}\cmidrule(lr){7-9}
   &(200,100)&$\mathrm{EA}_{\bb_{.1}}$ & 1.05(0.12)   & 1.05(0.12)   & 1.05(0.12)   & 0.70(0.01)   & 0.70(0.01)   & 0.70(0.01)  \\
   &&$\mathrm{EA}_{\bb}$  & 0.16(0.02)   & 0.16(0.02)   & 0.16(0.02)   & 0.15(1e-3) & 0.15(1e-3) & 0.15(1e-3) \\
             \hline
    \end{tabular}%
\end{table}%

\begin{table}[H]
    \begin{tabular}{rlrrrrrrr}
        \end{tabular}
\end{table}

\begin{figure}[H]
  \centering
  \includegraphics[width=12cm]{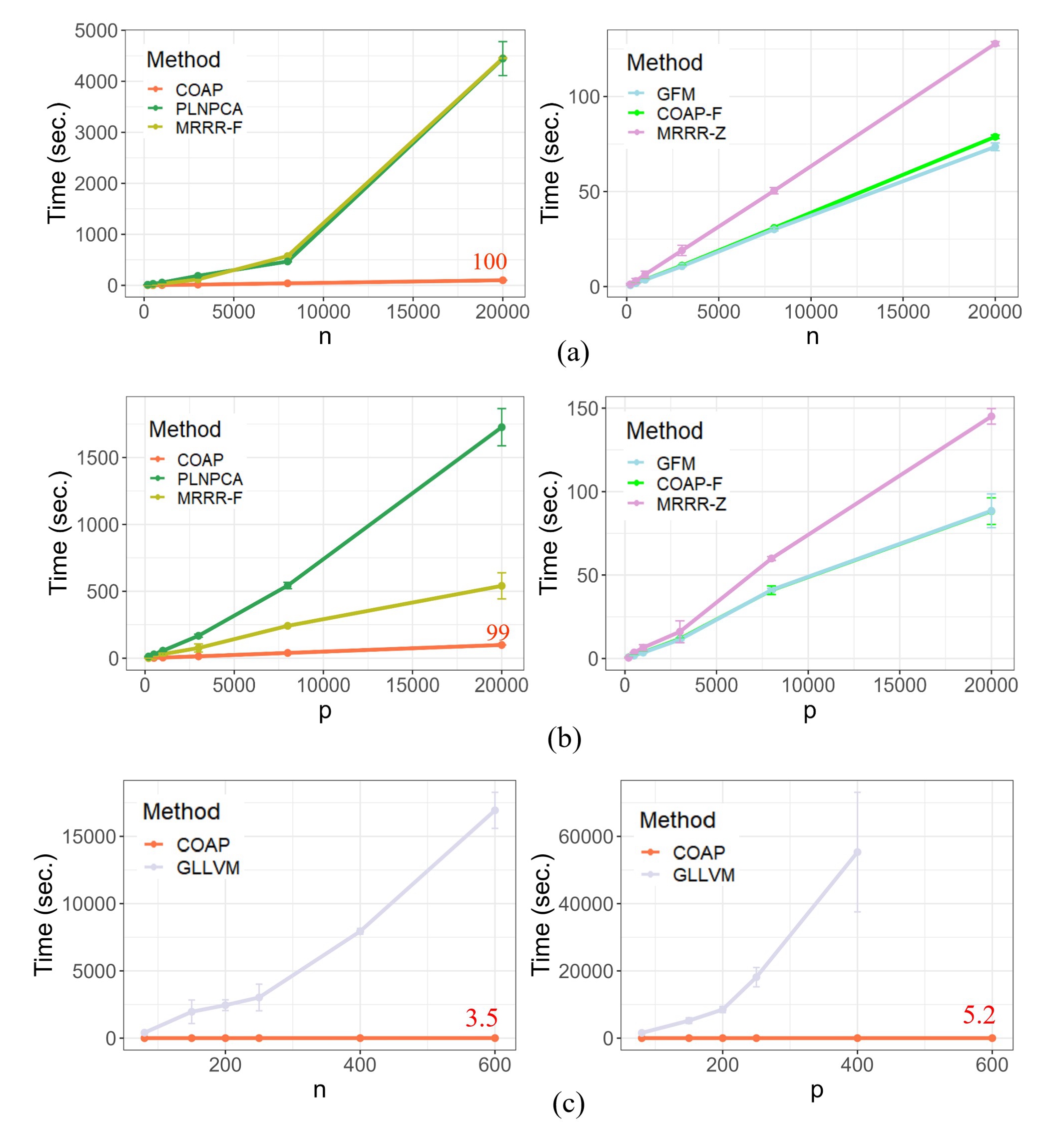}
  \caption{\red Comparison of computational time for COAP and other methods. (a) The running time versus sample size $n$ in case 1 with fixed $p=1000$; (b) the running time versus variable dimension $p$ in case 2 with fixed $n=1000$. COAP$_F$: the simplified model of COAP without covariates; (c) the running time of COAP and GLLVM versus $n$ while fixing $p=50$, and versus $p$ while fixing $n=80$, respectively. Note that GLLVM breaks down for the case with $(n=80, p=600)$. The values in red color represent the running time of COAP at the end point of the x-axis. }\label{fig:compTime}
\end{figure}

{\noindent\bf \underline{Scenario 5}}. An efficient algorithm is very important in analyzing large-scale high-dimensional data. In this scenario, we investigate the computational efficiency of COAP in comparison with {\red five other methods. The details regarding the methods compared and the data generation process have been deferred to Appendix D.2 in the Supporting Information.}  Figure \ref{fig:compTime} presents the average running time of COAP and other methods over ten repeats.  The results suggest that COAP is much more computationally efficient than  {\red PLNPCA, GLLVM and MRRR$_F$. Notably, GLLVM is the most inefficient method that spends 4.7 hours for the case with $(n=600,p=50)$ and $15.4$ hours for $(n=80, p=400)$ on average, and even breaks down for $(n=80, p=600)$, while COAP is significantly faster, with an average running time of 3.5 seconds for $(n=600, p=50)$ and 5.2 seconds for $(n=80, p=600)$.} Moreover, the simplified version COAP$_F$ runs faster than MRRR$_Z$ while being close to GFM.
\nvs

\nvs
\section{Real data analysis} \label{sec:real}
To demonstrate the utility of COAP, we applied it to the motivating data consisting of a {\red high-dimensional gene expression dataset (GEO accession number: GSM4732115), augmented with additional protein markers (GEO accession number: GSM4732116)} from a donor's peripheral blood mononuclear cells (PBMC)~\citep{mimitou2021scalable}. This PBMC dataset was collected by a multi-modal sequencing technology known as CITE-seq
~\citep{stoeckius2017simultaneous}, which can simultaneously measure the expression levels of tens of thousands of genes and hundreds of proteins. The PBMC data consist of 33,538 genes and 227 proteins for 3,480 cells after filtering the low-quality cells.  Following the guideline of CITE-seq sequencing data analysis~\citep{stoeckius2017simultaneous}, we first selected the top 2,000 highly variable genes of high quality. Our primary objective was to investigate the association between the expression levels of these 2,000 genes and 207 protein markers while accounting for unobserved latent factors.

\begin{figure}
  \centering
  \includegraphics[width=12cm]{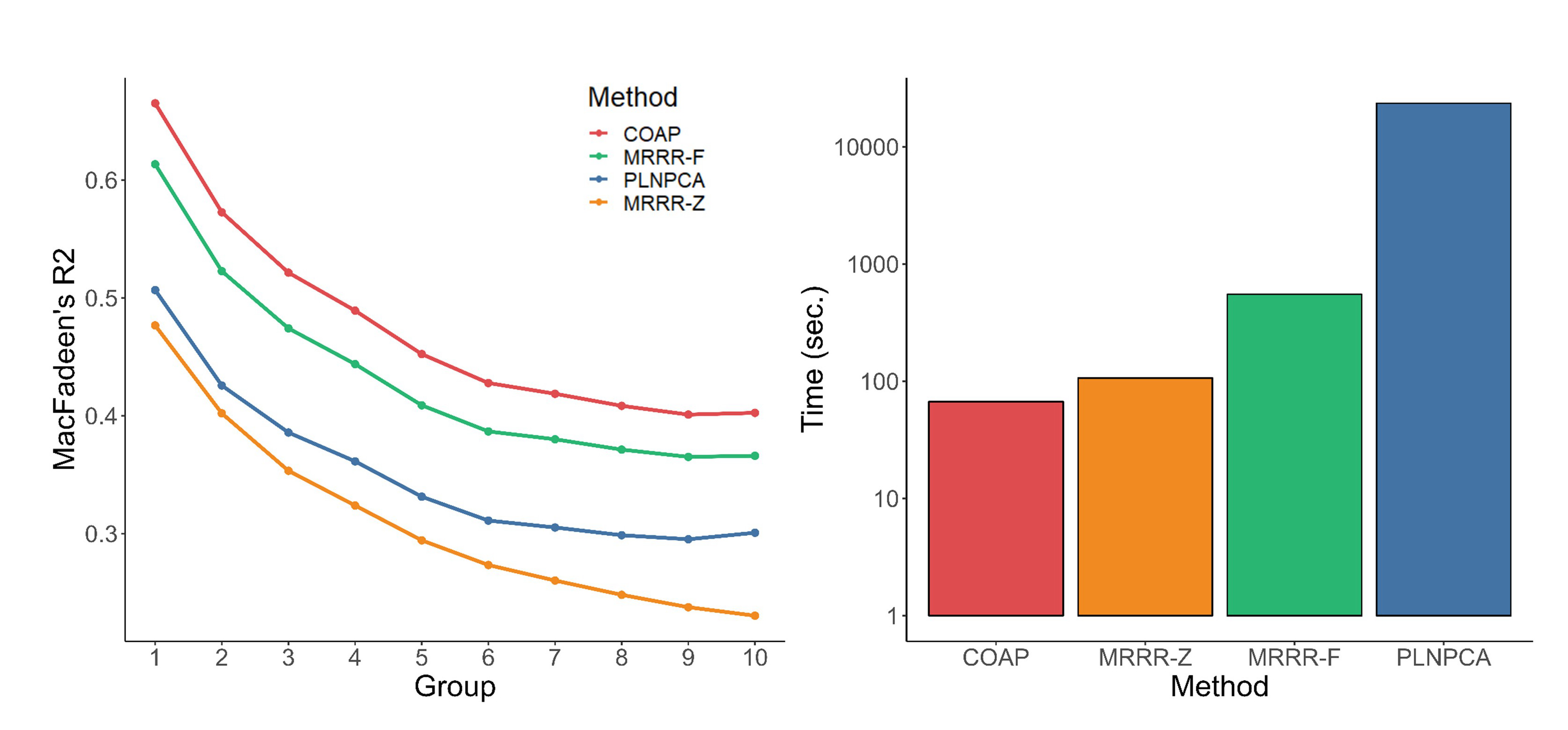}
  \caption{Comparison of feature extraction and computational time for COAP and other methods. Left panel: average MacFadden's $R^2$ of each group; right panel: running time of COAP and other methods. (Note: y-axis is in log scale).}\label{fig:compMacR2}
\end{figure}

Based on the SVR criterion, we selected the number of factors and the rank of $\bb$ as $\widehat q= 14$ and $\widehat r= 4$, respectively, by setting the upper bounds to 15 and 24, respectively. We then fit COAP using $\widehat q= 14$ and $\widehat r= 4$. In addition, we also implemented PLNPCA, MRRR$_Z$, and MRRR$_F$ as they can handle the high-dimensional data with additional covariates. For a fair comparison, we fixed the number of factors in PLNPCA and MRRR$_F$ as $\widehat q=14$ and set the rank in MRRR$_Z$ and MRRR$_F$ as $\widehat r = 4$. Since there is no ground truth for this data, we compared the performance of feature extraction for COAP and other methods by calculating an adjusted MacFadden's $R^2$ for each gene $x_{ij}$ $(j=1,\cdots, 2000)$. Specifically, for COAP, PLNPCA, and MRRR$_F$, we defined the extracted features as $\F = (\wh\H, \Z \wh\V_{\bb}) \in \mathbb{R}^{n\times (\widehat q+ \widehat r)}$, where $\wh \V_{\bb}\in R^{d\times \widehat r}$ is the rank-$\widehat r$ right singular matrix of $\wh\bb$. Since MRRR$_Z$ does not model the factor structure, we defined the extracted features as $\F=\Z \wh \V_{\bb}$. We then fitted a Poisson regression model between $\X_j$ and $\F$ for each gene $j$ and calculated the adjusted MacFadden's $R^2$ ($R^2_j$) for the fitted model~\citep{cameron1996r}. $R^2_j$ measures how much information of $\X_j$ is contained in $\F$, and a larger value implies better performance in feature extraction. We divided the genes into ten groups with 200 genes in each group and computed the mean of $R^2_j$ in each group. Figure \ref{fig:compMacR2} shows that COAP outperformed other methods in feature extraction across all ten groups. Furthermore, COAP exhibited high computational efficiency, {\red taking only 67 seconds}, whereas PLNPCA required 14,000 seconds (the right panel, Figure \ref{fig:compMacR2}).
\nvs
\vspace{-0.2in}

\begin{figure}
  \centering
  \includegraphics[width=14cm]{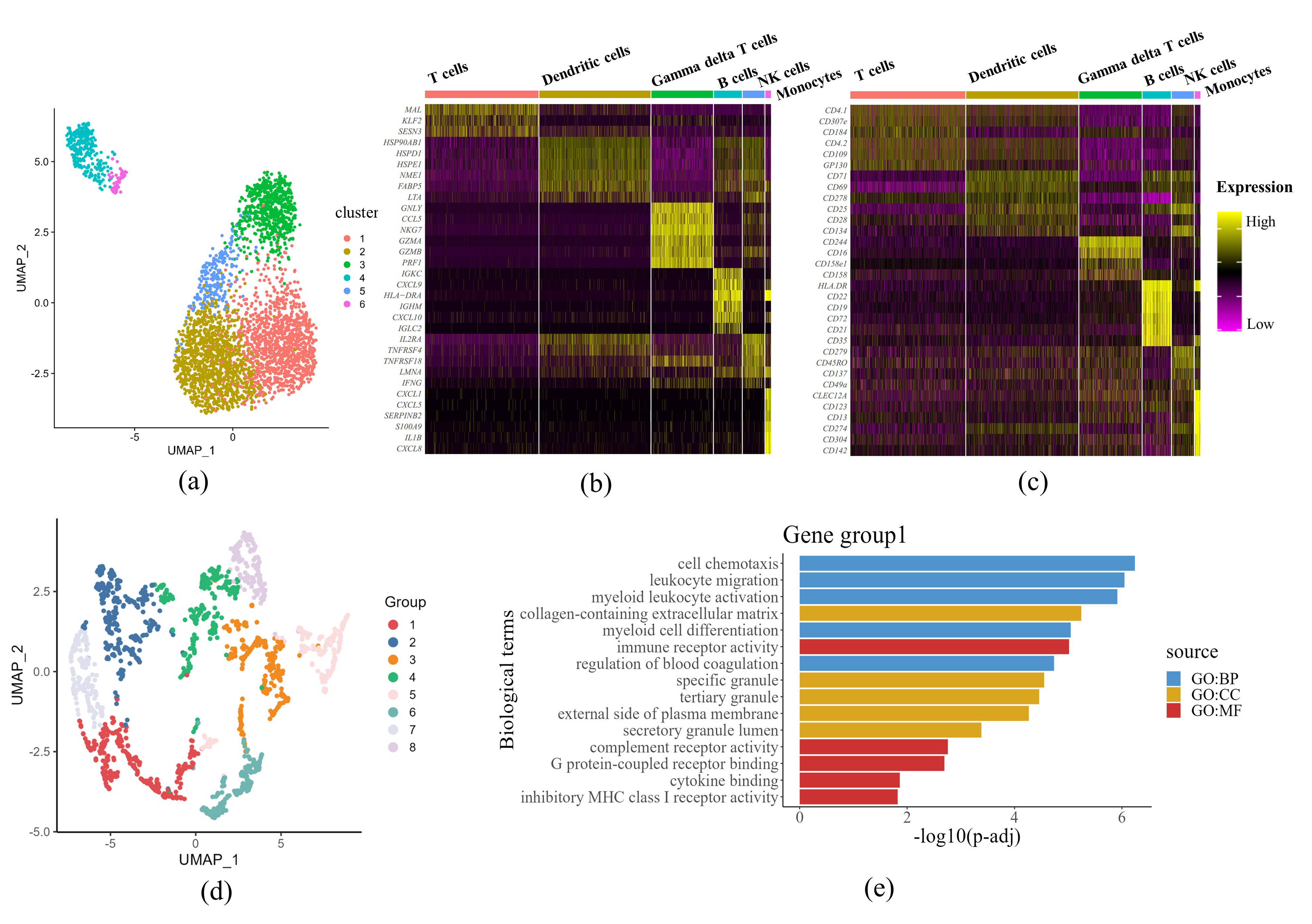}
  \caption{Downstream analysis of COAP results. (a) UMAP plot of six cell clusters identified by clustering on $\F$; (b) Heatmap of differentially expressed genes for
each cell type; (c) Heatmap of differentially expressed proteins for
each cell type; (d) UMAP plot of eight gene groups identified by clustering on $\U_{\bb}$; {(e) Top pathways
for the first group of genes}.}\label{fig:heatmapUMAP}
\end{figure}

\vspace{-0.1in}
The extracted features and estimated coefficient matrix obtained from COAP provide valuable benefits to downstream analysis tasks.  First, we use the extracted feature matrix $\F$ from COAP to cluster cells, followed by conducting differential expression analysis to find marker genes and proteins, and ultimately identifying cell types based on the identified marker genes. We perform the Louvain clustering method on $\F$, which is widely used in single-cell RNA sequencing data analysis~\citep{stuart2019comprehensive}, and identify six distinct cell clusters. By visualizing the six clusters on two-dimensional UMAP embeddings~\citep{becht2019dimensionality} extracted from $\F$ (Figure \ref{fig:heatmapUMAP}(a)), we observe good separation of different cell clusters. Using the detected clusters, we identify marker genes and proteins for each cluster using a Wilcoxon rank sum test, and subsequently classify the cell types for clusters 1 to 6 as T cells, dendritic cells, gamma delta T cells, B cells, natural killer (NK) cells, and monocytes, respectively, utilizing the PanglaoDB database (\url{https://panglaodb.se/}). We further visualize the expression levels of the top six marker genes (Figure \ref{fig:heatmapUMAP}(b)) and marker proteins (Figure \ref{fig:heatmapUMAP}(c)) for each cell type, which highlights the clear separation of differentially expressed genes and proteins across different cell types.

One of the most important features of COAP is its ability to group genes and proteins based on their association relationships, which can aid in exploring the functions of grouped genes or proteins. Let $\overline \bb \in \mathbb{R}^{2000\times 227}$ be the last 227 columns of $\wh\bb \in \mathbb{R}^{2000\times 228}$ obtained from COAP by dropping the intercepts. Using the SVD of $\overline \bb=\U_{\bb}  \bLambda_{\bb}  \V^{\trans}_{\bb}$, where $\U_{\bb}\in \mathbb{R}^{2000\times 4}$ represents the associations in the gene space and $\V_{\bb}\in \mathbb{R}^{227\times 4}$ represents the associations in the protein space, {\red we are able to cluster the genes into eight groups by applying Louvain clustering}  to $\U_{\bb}$. The groups are visualized in the two-dimensional UMAP embeddings extracted from $\U_{\bb}$ (Figure \ref{fig:heatmapUMAP}(d)), demonstrating good separation. To explore the functions of each group, we used the Gene Ontology database (\url{http://geneontology.org/}) and plotted the top five pathways of biological processes, molecular functions, and cellular composition for each group. Figure \ref{fig:heatmapUMAP}(e)  shows that the first group of genes is primarily associated with biological processes related to {\red the myeloid leukocyte activation, myeloid cell differentiation and immune receptor activity.} The top pathways for other groups are shown in Figure S3 of Supporting Information. Finally, we show that $\U_{\bb}$ and $\V_{\bb}$ allow us to identify the top genes and proteins with the strongest associations. For comprehensive findings in this regard, refer to Appendix D.3 in the Supporting Information.
\nvs

\vspace{-0.6in}
\section{Discussion}\label{sec:dis}
We have developed a novel statistical model, called COAP, to analyze high-dimensional overdispersed count data with additional high-diemensional covariates. This model is particularly useful in scenarios where the sample size ($n$), count variable dimension ($p$), and covariates dimension ($d$) are all large.
We have theoretically studied the computational identifiability conditions of the model and proposed a computationally efficient VEM algorithm.
We also developed a singular value ratio based method to determine the number of factors and rank of the coefficient matrix. In numerical experiments, we demonstrated that COAP outperforms existing methods in terms of both estimation accuracy and computation time, making it an attractive alternative for analyzing large-scale count data. Furthermore, our PBMC CITE-seq study demonstrated that COAP is a powerful tool for understanding the underlying structure of complex count data. We believe that COAP has the potential to facilitate important discoveries not only in genomics but also in other fields.

{\red A straightforward extension for COAP involves managing multiple modalities that have various variable types in each modality and additional covariates. As an illustration, SNARE-seq~\citep{zhu2020single} is a sequencing technology for transcriptomics and epigenomics that can measure RNA expressions and chromatin accessibility simultaneously. These are often represented by high-dimensional count and binary variables, respectively. Moreover, within the COAP framework, a new and exciting avenue of research is the exploration of spatial genomics data~\citep{liu2023high}, where the spatial location of each cell $i$ is measured in addition to the expression data of multi-omics.}

\nvs
\bibliographystyle{biom}

\bibliography{ref_library}

\end{document}